\documentclass[12pt]{article}
\usepackage{amssymb, amsmath, amsopn, amsthm}

\usepackage{bm,amsfonts,array,calc,rotating}
\usepackage{epsfig}
\usepackage{graphicx}
\usepackage{color}
\usepackage{slashed}
\usepackage[colorlinks=false]{hyperref}

\global\def\draftcontrol{0}

   \def\versionno{Higher Derivative Brane Couplings from T-Duality}

\catcode`\@=11

\expandafter\ifx\csname
draftcontrol\endcsname\relax\global\def\draftcontrol{0} \fi

{\count255=\time\divide\count255 by 60
\xdef\hourmin{\number\count255} \multiply\count255
by-60\advance\count255 by\time
\xdef\hourmin{\hourmin:\ifnum\count255<10 0\fi\the\count255}}
\def\draftdate{\number\month/\number\day/\number\year\ \ \ \hourmin }

\newcommand\makepapertitle{\par

  \begingroup
    \renewcommand\thefootnote{\@fnsymbol\c@footnote}%
    \def\@makefnmark{\rlap{\@textsuperscript{\normalfont\@thefnmark}}}%
    \long\def\@makefntext##1{\parindent 1em\noindent
            \hb@xt@1.8em{%
                \hss\@textsuperscript{\normalfont\@thefnmark}}##1}%
     \newpage
     \global\@topnum\z@   
     \@makepapertitle
     \thispagestyle{empty}\@thanks
  \endgroup
  \setcounter{footnote}{0}%
  \global\let\thanks\relax
  \global\let\makepapertitle\relax
  \global\let\@makepapertitle\relax
  \global\let\@thanks\@empty
  \global\let\@author\@empty
  \global\let\@date\@empty
  \global\let\@title\@empty
  \global\let\title\relax
  \global\let\author\relax
  \global\let\date\relax
  \global\let\and\relax
  \def\version{\let\version\@version\@gobble}
}
\def\@makepapertitle{%
  \newpage
   \ifnum\draftcontrol=1{}
   \version\versionno
   \vskip 5em%
   \else
   \hfill\hbox to 3cm {\parbox{4cm}{\@pubnum}\hss}%
   \vskip 5em%
   \fi
   \begin{center}%
   \let \footnote \thanks
      {\hskip -0\textwidth \hbox to 1\textwidth%
        {\centerline{\Large\bf{\noindent\@title}}}}%
     \vskip 2em%
     {\normalsize
       \lineskip .5em%
       \begin{tabular}[t]{c}%
         \@author
       \end{tabular}\par}%
     \vskip 1em%
     {\@bstract}%
     \end{center}%
     \vfill
     \@date%
     \vskip 1.5em%
     \noindent
     \rule{12em}{.02em}\par\noindent
     \@email%
   \par
}

\gdef\@pubnum{}
\def\pubnum#1{%
  \gdef\@pubnum{#1}}

\gdef\@bstract{}
\def\Abstract#1{%
  \gdef\@bstract{%
   \parbox{\textwidth-0pc}{%
   \centerline{\bf Abstract}\penalty1000
   \noindent
   \renewcommand\baselinestretch{1.0}
   {#1}}}
}

\gdef\@email{}
\def\email#1{%
   \gdef\@email{%
  {\small Email: {\tt #1}}}
}

\def\ps@paper{\let\@mkboth\@gobbletwo%
     \ifnum\draftcontrol=1
        \def\@oddfoot{\hbox to \textwidth{\tiny \versionno \hfil\tiny\draftdate}%
        \hskip -\textwidth \hbox to \textwidth{\hfil\rm\thepage\hfil}}%
     \else\def\@oddfoot{\hbox to \textwidth{\hfil\rm\thepage\hfil}}
     \fi
     \let\@evenfoot\@oddfoot
}

\def\body{\clearpage
          \pagestyle{paper}
        }
\newenvironment{acknowledgments}{%
\vskip 3.25ex
\noindent {\bf Acknowledgments}
}

\def\@version#1{\ifnum\draftcontrol=1
\typeout{}\typeout{#1}\typeout{}
\vskip3mm\centerline{\hbox{\fbox{\normalsize{\tt DRAFT -- #1 -- }
                   {\draftdate}}}}\vskip3mm
\fi}
\let\version\@version
\long\def\eqlabel#1{\ifnum\draftcontrol=1
                    \tag@false  
                    \tag*{(\theequation) \hbox to -0.2cm{\hspace{0cm}\small{#1}\hss}}
                    \refstepcounter{equation}
                    \edef\@currentlabel{\theequation}
                    \ltx@label{#1}          
                    \else
                    \label{#1}
                    \fi
                    }
\let\st@bibitem\@bibitem
\let\st@lbibitem\@lbibitem
\ifnum\draftcontrol=1
  \def\@bibitem#1{%
    \st@bibitem{#1}\a@@label{#1}\ignorespaces}
  \def\@lbibitem[#1]#2{%
    \st@lbibitem[#1]{#2}\a@@label{#2}\ignorespaces}
  \def\a@@label#1{%
    \gdef\a@lab{\smash{\normalfont\small#1}}
    \ifvmode
      \if@inlabel
        \global\setbox\@labels\hbox{%
          \llap{\a@lab\let\a@lab\relax
                \kern\@totalleftmargin\kern\marginparsep}%
          \box\@labels}%
      \fi
    \fi}
\fi

\renewcommand\baselinestretch{1.25}
\renewcommand\section{\@startsection {section}{1}{\z@}%
                                   {-3.5ex \@plus -1ex \@minus -.2ex}%
                                   {2.3ex \@plus.2ex}%
                                   {\normalfont\large\bfseries}}
\renewcommand\subsection{\@startsection{subsection}{2}{\z@}%
                                   {-3.25ex\@plus -1ex \@minus -.2ex}%
                                   {1.5ex \@plus .2ex}%
                                   {\normalfont\normalsize\bfseries}}
\renewcommand\subsubsection{\@startsection{subsubsection}{3}{\z@}%
                                   {-3.25ex\@plus -1ex \@minus -.2ex}%
                                   {1.5ex \@plus .2ex}%
                                   {\normalfont\normalsize\it}}
\renewcommand\paragraph{\@startsection{paragraph}{4}{\z@}%
                                   {-3.25ex\@plus -1ex \@minus -.2ex}%
                                   {1.5ex \@plus .2ex}%
                                   {\normalfont\normalsize\bf}}
\renewcommand\subparagraph{\@startsection{subparagraph}{5}{\z@}%
                                   {-1.25ex\@plus -1ex \@minus -.2ex}%
                                   {0ex \@plus .2ex}%
                                   {\normalfont\normalsize\it}}

\numberwithin{equation}{section}

\long\def\@makecaption#1#2{%
  \vskip\abovecaptionskip
  \sbox\@tempboxa{{\bf #1:} #2}%
  \ifdim \wd\@tempboxa >\hsize
    {\small\bf #1:} {\small #2}\par
  \else
    \global \@minipagefalse
    \hb@xt@\hsize{\hfil\box\@tempboxa\hfil}%
  \fi
  \vskip\belowcaptionskip}

\renewcommand*\l@section[2]{%
  \ifnum \c@tocdepth >\z@
    \addpenalty\@secpenalty
    \addvspace{.5em \@plus\p@}%
    \setlength\@tempdima{1.5em}%
    \begingroup
      \parindent \z@ \rightskip \@pnumwidth
      \parfillskip -\@pnumwidth
      \leavevmode \bfseries
      \advance\leftskip\@tempdima
      \hskip -\leftskip
      #1\nobreak\hfil \nobreak\hb@xt@\@pnumwidth{\hss #2}\par
    \endgroup
  \fi}
\renewcommand*\l@subsection{\addvspace{.0em \@plus\p@}\@dottedtocline{2}{1.5em}{2.3em}}
\renewcommand*\l@subsubsection{\addvspace{-.2em \@plus\p@}\@dottedtocline{3}{3.8em}{3.2em}}

\definecolor{refcol}{rgb}{0.2,0.2,0.8}
\definecolor{eqcol}{rgb}{.6,0,0}
\definecolor{purple}{cmyk}{0,1,0,0}

\gdef\@citecolor{refcol} \gdef\@linkcolor{eqcol}
\def\colorlinkspurple{\gdef\@urlcolor{purple}}
\def\colorlinksblue{\gdef\@urlcolor{blue}}
\def\colorlinksred{\gdef\@urlcolor{red}}

\def\revise#1       {\raisebox{-0em}{\rule{3pt}{1em}}%
                     \marginpar{\raisebox{.5em}{\vrule width3pt\
                     \vrule width0pt height 0pt depth0.5em
                     \hbox to 0cm{\hspace{0cm}{%
                     \parbox[t]{4em}{\raggedright\footnotesize{#1}}}\hss}}}}

\def\a{\alpha}
\def\b{\beta}
\def\g{\gamma}
\def\d{\delta}

\def\th{\theta}

\def\m{\mu}
\def\n{\nu}

\def\s{\sigma}

\def\G{\Gamma}

\def\L{\Lambda}
\def\O{\Omega}

\def\p{\partial}

\def\p{\partial}

\def\R{{\rm I\!R }}

\def\inbar{\,\vrule height1.5ex width.4pt depth0pt}
\def\IC{\relax\hbox{$\inbar\kern-.3em{\rm C}$}}
\def\IN{\relax{\rm I\kern-.18em N}}
\def\IR{\relax{\rm I\kern-.18em R}}
\font\cmss=cmss10 \font\cmsss=cmss10 at 7pt
\def\IZ{\relax\ifmmode\mathchoice
{\hbox{\cmss Z\kern-.4em Z}}{\hbox{\cmss Z\kern-.4em Z}}
{\lower.9pt\hbox{\cmsss Z\kern-.4em Z}} {\lower1.2pt\hbox{\cmsss
Z\kern-.4em Z}}\else{\cmss Z\kern-.4em Z}\fi}

\def\a{\alpha}
\def\b{\beta}
\def\be{\begin{equation}}
\def\beq{\begin{eqnarray}}
\def\g{\gamma}
\def\d{\delta}
\def\ee{\end{equation}}
\def\eeq{\end{eqnarray}}

\def\th{\theta}

\def\m{\mu}
\def\n{\nu}

\def\s{\sigma}

\def\G{\Gamma}

\def\L{\Lambda}
\def\O{\Omega}

\def\p{\partial}

\def\p{\partial}

\def\b{\beta}
\def\g{\gamma}

\def\a{\alpha}
\def\G{\Gamma }

\def\e{\varepsilon}

\def\inbar{\,\vrule height1.5ex width.4pt depth0pt}

\def\zbar{\bar z}
\def\pa{\partial}

\def\ee           {{\it e}}

\def\tr           {{\rm tr}}
\def\Tr        {{\rm Tr}}

\def\sqr#1#2{{\vcenter{\vbox{\hrule height.#2pt
 \hbox{\vrule width.#2pt height#1pt \kern#1pt
 \vrule width.#2pt}\hrule height.#2pt}}}}

\newcommand{\lp}{\left(}
\newcommand{\rp}{\right)}
\newcommand{\ls}{\left[}
\newcommand{\rs}{\right]}
\newcommand{\non}{\nonumber}
\newcommand{\hlf}{\frac{1}{2}}
\renewcommand{\be}{\begin{equation}}
\renewcommand{\ee}{\end{equation}}
\newcommand{\w}{\wedge}

\catcode`\@=12

\begin{document}

\pubnum{MIFP-10-29}

\title{Higher Derivative Brane Couplings from T-Duality}

\date{September 7, 2010}

\author{\\[.5cm]Katrin Becker, Guangyu Guo and Daniel Robbins\
\\[.2cm] \it Department of Physics, Texas A\&M University, \\ \it College Station, TX 77843, USA\\ [1.5cm]}

\Abstract{The Wess-Zumino coupling on D-branes in string theory is known to receive higher derivative corrections which couple the Ramond-Ramond potential to terms involving the square of the spacetime curvature tensor.  Consistency with T-duality implies that the branes should also have four-derivative couplings that involve the NS-NS $B$-field.  We use T-duality to predict some of these couplings.  We then confirm these results with string worldsheet computations by evaluating disc amplitudes with insertions of one R-R and two NS-NS vertex operators.}

 \email{kbecker, guangyu, robbins@physics.tamu.edu}

\makepapertitle

\body


\vskip 1em

\newpage

\section{Introduction}

In order to make progress towards connecting string theory to real
world physics, it is crucial to understand the vacua of the theory -
what ingredients can be used to construct vacua and what consistency
conditions constrain the possible ways of assembling these
ingredients.  Of particular interest in vacuum construction are
D-branes (localized objects on which strings can end) and fluxes
(various higher-dimensional analogs of magnetic fields).

Fluxes have become a very important ingredient in constructing semi-realistic vacua, as they provide a straightforward mechanism to give masses to the many scalar fields which describe the geometry of the internal manifold.  In type II string theory, for example, there are fluxes corresponding to the NS-NS $B$-field, whose field strength is a three-form $H_3$, and various R-R $p$-form fields $C^{(p)}$, with field strengths $F^{(p+1)}$.  The term {\it{fluxes}} is most commonly used to describe the discrete topological parameters that come from integrating these field-strengths over cycles in the internal manifold.  In this paper, however, we are always working locally, and so we won't be dealing directly with these topological quantities.  We do, however, deal with situations where some of these potentials (especially $B$) have non-vanishing derivatives, and so we expect our results to be important especially in the presence of fluxes.

Another important set of ingredients are D-branes - non-perturbative excitations in the theory which are localized to a sub-manifold of the ten-dimensional spacetime.  The D-branes that we will be considering carry R-R charge (and hence are stable to decay), and there are many degrees of freedom localized to their worldvolumes.  These localized degrees of freedom are one of the reasons that D-branes are so attractive in vacuum building, as they can include chiral matter and non-abelian gauge groups.

D-branes can also be very important in finding consistent compactifications, as they can sometimes be needed to satisfy an important
class of consistency conditions known as tadpole equations.  In particular, tadpole equations can impose
discrete topological constraints on the number and type of D-branes
and quantized fluxes.

For instance in type IIB, the equation of motion for $C^{(4)}$ wrapping the directions of Minkowski space is an internal closed six-form which gets contributions both from fluxes (terms proportional to $F^{(3)}\w H_3$) and from delta-function forms corresponding to localized sources such as D3-branes and O3-planes, and can also receive contributions from higher-derivative corrections to the action.  If the six-form is not exact, then there can be a topological obstruction to solving the tadpole equation, and the compactification would be inconsistent.  In fact, it turns out that in some examples of this sort (as well as in some other contexts), there may be no way to solve the tadpole constraint at leading order in a momentum expansion.  Higher derivative corrections must then be included that often change the global structure drastically - either by allowing the existence of solutions, or perhaps by spoiling the consistency of solutions that otherwise appeared to be fine.  For this reason, it is crucial to understand these corrections and their global properties.

The IIB case mentioned above is an excellent example.  The local
tadpole equation gets modified by higher derivative terms which,
when integrated over the internal space, gives a definite
topological contribution, proportional to the Euler number of the
auxiliary Calabi-Yau four-fold in F-theory.  In a limit in which the
compactification is well described by type IIB with O7-planes and
D7-branes, the higher derivative corrections have precisely the form
of a four-derivative correction to the action localized at these O7
and D7 sources.  The leading piece of the action from the D7-branes
is a Wess-Zumino action \be
S_{WZ}=T_7\int_{D7}Ce^{B+2\pi\a'F}|_{\mathrm{8-form}}. \ee At the
O7-planes we have something similar, but the pull-back of $B$
vanishes, there's no gauge field, and the numerical coefficient is
different.  These actions do get corrections depending on
derivatives of the bulk metric~\cite{
Bershadsky:1995,Green:1996dd,Cheung:1997az,Dsagupta:1997,Minasian:1997mm},
\be
Ce^{B+2\pi\a'F}|_{\mathrm{8-form}}+\frac{\pi^2\lp\a'\rp^2}{24}Ce^{B+2\pi\a'F}|_{\mathrm{4-form}}\w\lp\tr
R_T\w R_T-\tr R_N\w R_N\rp+\mathcal{O}((\a')^4), \ee where $R_T$ and
$R_N$ are curvatures (we will explain our notation more fully in
section \ref{sec:Predictions}). As emphasized above, these
higher-derivative terms really must be included in order to
accurately judge the consistency of a given solution. But these
terms are not the end of the story.  They provide a particular set
of four-derivative couplings on the brane between the bulk spacetime
metric and the R-R potential which contribute crucially to the
$C^{(4)}$ tadpole.  However, there can be many other couplings
between $C^{(4)}$ and bulk NS-NS fields at this same derivative
order.  Indeed, by using T-duality, one can deduce some more
couplings which involve derivatives of $B$-fields, or will involve
R-R fields of different degree, etc.  It is not clear that these
couplings will necessarily lead to new topological restrictions, but
in some contexts they might, and they will certainly modify the
local tadpole equation.  Similar couplings have been obtained via
U-duality in M-theory and string theory
in~\cite{sethitalks,mcoristsethitoappear}, where they have been used
to avoid no-go theorems in IIA and M-theory flux compactifications.
Clearly, these issues need to be examined more closely than they
have been.

\subsection{Summary of Results}

In this paper we start with some of the known corrections to the Wess-Zumino term in the action of a D$p$-brane,
\be
\label{eq:OriginalWZAction}
S_{WZ\,\mathrm{original}}=T_p\frac{\pi^2(\a')^2}{24}\int_{Dp}Ce^B\lp\tr R_T\w R_T-\tr R_N\w R_N\rp.
\ee

By analyzing the conditions which are imposed by consistency with T-duality, we will show that the action must contain these terms as well as several others at this order,
\begin{multline}
\label{eq:FixedCouplings}
S_{WZ}\supset T_p\frac{\pi^2\lp\a'\rp^2}{24}\int_{Dp} dx^{a_1}\w\cdots\w dx^{a_{p+1}} \\
\left\{\hlf\frac{1}{\lp p-3\rp !}C^{(p-3)}_{a_1\cdots a_{p-3}}\lp -2\p_{a_{p-2}}^{\hphantom{a_{p-2}}[b}h_{a_{p-1}}^{\hphantom{a_{p-1}}c]}\p_{a_pb}h_{a_{p+1}c}+2\p_{a_{p-2}}^{\hphantom{a_{p-2}}[j}h_{a_{p-1}}^{\hphantom{a_{p-1}}k]}\p_{a_pj}h_{a_{p+1}k}\right.\right. \\
\left.\left. -\p_{a_{p-2}}^{\hphantom{a_{p-2}}b}B_{a_{p-1}}^{\hphantom{a_{p-1}}j}\p_{a_pb}B_{a_{p+1}j}+\p_{a_{p-2}}^{\hphantom{a_{p-2}}j}B_{a_{p-1}}^{\hphantom{a_{p-1}}b}\p_{a_pj}B_{a_{p+1}b}\rp\right. \\
\left.+\frac{1}{\lp p-2\rp !}C^{(p-1)}_{a_1\cdots a_{p-2}i}\lp 2\p_{a_{p-1}}^{\hphantom{a_{p-1}}[b}h_{a_p}^{\hphantom{a_p}c]}\p_{a_{p+1}b}B^i_{\hphantom{i}c}-2\p_{a_{p-1}}^{\hphantom{a_{p-1}}[j}h_{a_p}^{\hphantom{a_p}k]}\p_{a_{p+1}j}B^i_{\hphantom{i}k}\right.\right. \\
\left.\left.+\p_{a_{p-1}}^{\hphantom{a_{p-1}}b}h^{ij}\p_{a_pb}B_{a_{p+1}j}-\p_{a_{p-1}}^{\hphantom{a_{p-1}}j}h^{ib}\p_{a_pj}B_{a_{p+1}b}\rp\right. \\
\left.+\hlf\frac{1}{\lp p-1\rp !}C^{(p+1)}_{a_1\cdots a_{p-1}i_1i_2}\lp -\p_{a_p}^{\hphantom{a_p}b}h^{i_1j}\p_{a_{p+1}b}h^{i_2}_{\hphantom{i_2}j}+\p_{a_p}^{\hphantom{a_p}j}h^{i_1b}\p_{a_{p+1}j}h^{i_2}_{\hphantom{i_2}b}\right.\right. \\
\left.\left. -2\p_{a_p}^{\hphantom{a_p}b}B^{i_1c}\p_{a_{p+1}[b}B^{i_2}_{\hphantom{i_2}c]}+2\p_{a_p}^{\hphantom{a_p}j}B^{i_1k}\p_{a_{p+1}[j}B^{i_2}_{\hphantom{i_2}k]}\rp\right\}.
\end{multline}
Here we have expanded around a D-brane with the usual static gauge embedding in a flat background with no $B$-field.  Indices from the beginning of the alphabet run over directions along the worldvolume of the D-brane, while indices from the middle of the alphabet run over the transverse directions.  We have included metric fluctuations, $g_{\m\n}=\eta_{\m\n}+h_{\m\n}$, and $B$-field fluctuations, $B_{\m\n}$, as well as fluctuations of the R-R potentials of degrees $(p-3)$, $(p-1)$, and $(p+1)$, and we have only worked to first order in R-R fluctuations and quadratic order in NS-NS fluctuations.  The first line inside the curly braces comes from expanding the known couplings (\ref{eq:OriginalWZAction}) to this order in fluctuations.  The remaining five lines are new couplings.

To check these results, we will compute disc amplitudes involving the insertion of one closed string R-R vertex operator and two NS-NS vertex operators.  The results of this computation will agree with (\ref{eq:FixedCouplings}) up to an overall normalization, which can in turn be fixed by comparing with (\ref{eq:OriginalWZAction}).

In section \ref{sec:Predictions} we will use spacetime T-duality to argue for the presence of these additional terms, and we will in fact use the Buscher rules to compute several terms which must be present, eventually arriving at (\ref{eq:FixedCouplings}), which is the key result of this section.  In section \ref{sec:DiscAmplitudes} we will confirm these predictions by doing three-point disc amplitude computations, involving one R-R field and two NS-NS fields in the presence of a D-brane.  There are subtleties in the computation for general couplings in this type of amplitude which would require careful addition of boundary terms to settle - without adding the correct boundary terms, one finds disagreements when performing the computation in different pictures, for example.  However, we are fortunate that the particular terms which we predicted from T-duality do not require these boundary terms, so we may proceed with the somewhat naive computation.  In order to bolster our assertion that extra boundary terms are not needed, we perform the computation in several different pictures and confirm that in each case the results agree with the other cases and with the spacetime T-duality prediction.  We conclude in section \ref{sec:Discussion}.

\section{Predictions from T-Duality}
\label{sec:Predictions}

\subsection{Buscher rules}

The low-energy bosonic spectrum of type II closed string theory includes the metric $g_{\m\n}$, the two-form potential $B_{\m\n}$, and the dilaton $\Phi$ from the NS-NS sector, and $p$-form potentials $C^{(p)}$ from the R-R sector, where $p$ is odd for IIA or even for IIB.

In backgrounds which include a $U(1)$ isometry, string theory appears to enjoy a duality, called T-duality, relating one background which solves the equations of motion to another.  Pick coordinates such that the isometry corresponds to translation in one coordinate, $y$, and let the remaining coordinates be labeled by indices $\m$, $\n$, etc.  Then the explicit T-duality transformations for the NS-NS fields are given by~\cite{Buscher:1987sk}
\be
g'_{yy}=\frac{1}{g_{yy}},\qquad g'_{\m y}=\frac{B_{\m y}}{g_{yy}},\qquad g'_{\m\n}=g_{\m\n}-\frac{g_{\m y}g_{\n y}-B_{\m y}B_{\n y}}{g_{yy}},\non
\ee
\be
\label{eq:NSNSBuschers}
B'_{\m y}=\frac{g_{\m y}}{g_{yy}},\qquad B'_{\m\n}=B_{\m\n}-\frac{B_{\m y}g_{\n y}-g_{\m y}B_{\n y}}{g_{yy}},\qquad\Phi'=\Phi-\hlf\ln g_{yy},
\ee
and for the R-R potentials we have~\cite{Bergshoeff:1995as}
\beq
\label{eq:BuscherRR}
C^{(p)\prime}_{\m_1\cdots\m_{p-1}y} &=& C^{(p-1)}_{\m_1\cdots\m_{p-1}}-\lp p-1\rp\frac{C^{(p-1)}_{[\m_1\cdots\m_{p-2}|y|}g_{\m_{p-1}]y}}{g_{yy}},\\
C^{(p)\prime}_{\m_1\cdots\m_p} &=& C^{(p+1)}_{\m_1\cdots\m_py}+pC^{(p-1)}_{[\m_1\cdots\m_{p-1}}B_{\m_p]y}+p\lp p-1\rp\frac{C^{(p-1)}_{[\m_1\cdots\m_{p-2}|y|}B_{\m_{p-1}|y|}g_{\m_p]y}}{g_{yy}}.\non
\eeq


Under this duality, the type IIA and type IIB supergravity actions are mapped into each other, and in fact the action for the NS-NS sector fields is invariant under T-duality.

\subsection{Using T-duality to construct or constrain actions}

Suppose that we didn't actually know the two-derivative action for NS-NS sector fields, but knew only that it was invariant under diffeomorphisms and $B$-field gauge transformations.  In this case there are four possible terms we could write down in the Lagrangian,
\be
f_1(\Phi)\sqrt{-g}R,\qquad f_2(\Phi)\sqrt{-g}H^2,\qquad f_3(\Phi)\sqrt{-g}\nabla^2\Phi,\qquad f_4(\Phi)\sqrt{-g}\lp\nabla\Phi\rp^2,
\ee
where the $f_i$ are arbitrary functions of $\Phi$.  Note that one combination of these would be a total derivative, but if we continue to work at the level of Lagrangians, we can keep all four terms.  If we also know that the Lagrangian was invariant under the Buscher rules above, then we can actually fix the action up to an overall constant.  We would do this by assuming a background with a $U(1)$ isometry, evaluating each of the terms above in that situation, and demanding that the result be invariant.  One finds the invariant combination
\be
\mathcal{L}\supset \mathcal{N}e^{-2\Phi}\sqrt{-g}\lp R-\frac{1}{12}H^2+4\nabla^2\Phi-4\lp\nabla\Phi\rp^2\rp,
\ee
with $\mathcal{N}$ an arbitrary constant\footnote{One can compare this result with equation (1.10) of \cite{Hohm:2010jy}, which is obtained by slightly different reasoning.}.  If we knew the coefficient of one of the terms, like the Einstein-Hilbert term, then the other terms are determined.  In this way, T-duality can be used to fix the form of the action.

T-duality is also a useful guide in the presence of D-branes, converting a brane which wraps the direction of the $U(1)$ isometry into one which is localized at a point in the circle direction\footnote{In this discussion, we are referring to probe branes, not to branes or stacks of branes that backreact on the geometry.  A supergravity solution corresponding to a stack of branes wrapping a circle isometry with backreaction taken into account is converted, by T-duality, into a solution where a stack of lower-dimensional branes are smeared along the circle direction.  Instead, we are typically interested in only a single brane which is localized, not smeared.}.  T-duality should map the actions on such dual pairs of branes into one another.  In this paper we will be focused on the Wess-Zumino part of the D-brane action, its higher derivative corrections, and terms related to it by T-duality.  These terms can be written as
\be
T_p\int_{Dp}\mathcal{L}_{WZ}^{(p+1)},
\ee
where $T_p$ is the tension of the D-brane and $\mathcal{L}_{WZ}^{(p+1)}$ is a $(p+1)$-form on the worldvolume of the D-brane.  A naive guess for the zero-derivative piece of this action would be $\mathcal{L}_{WZ}^{(p+1)}=C^{(p+1)}$, but it turns out that this is inconsistent with T-duality.  Indeed, the requirement of consistency with T-duality is equivalent to demanding (we use a prime to indicate that the expression should be transformed by the Buscher rules (\ref{eq:NSNSBuschers}) and (\ref{eq:BuscherRR}))
\be
\label{eq:WZTDuality}
\mathcal{L}_{WZ\ \m_1\cdots\m_{p+1}}^{(p+1)\prime}=\mathcal{L}_{WZ\ \m_1\cdots\m_{p+1}y}^{(p+2)},\qquad\mathcal{L}_{WZ\ \m_1\cdots\m_py}^{(p+1)\prime}=\mathcal{L}_{WZ\ \m_1\cdots\m_p}^{(p)},
\ee
which is not satisfied by $C^{(p+1)}$ because of the non-linear pieces in the transformation rules (\ref{eq:BuscherRR}).  Rather, we should proceed as before and write down the possible terms which can appear, evaluate them in a circle isometry ansatz, and impose T-duality.  Doing so, we arrive at the T-duality completion of this naive term,
\be
\label{eq:LowestOrderWZ}
\mathcal{L}_{WZ}^{(p+1)}=Ce^B|_{(p+1)-\mathrm{form}},
\ee
where $C$ is a formal sum of R-R potentials and
\be
e^B=1+B+\hlf B\w B+\cdots.
\ee
It is not hard to see that (considered as forms in the ten-dimensional spacetime) the expression (\ref{eq:LowestOrderWZ}) satisfies (\ref{eq:WZTDuality}).

Thus, if one knew about T-duality, and knew that we expected at least a term in the Lagrangian like $\int_{Dp}C^{(p+1)}$, then we could deduce that it must be part of a larger ``T-duality invariant", $\int_{Dp}Ce^B$, where the $(p+1)$-form integrand here is understood to be pulled back to the worldvolume of the D$p$-brane.  Of course, if we also considered invariance under $B$-field gauge transformations, then we would be lead to introduce more terms, so that the final result was
\be
S_{WZ}^{(0)}=T_p\int_{Dp}Ce^{B+2\pi\a'F},
\ee
where $F=dA$ is the field strength of the worldvolume gauge field which transforms under $B$-field gauge transformations $B\rightarrow B+d\L$ as $A\rightarrow A-\L/(2\pi\a')$.

In most of what follows we will set the gauge field to zero, though of course the eventual task of constructing a full non-linear action will require its inclusion, along with many other terms that we have not written down, in order to satisfy $B$-field gauge invariance~\cite{inprogress}.

\subsection{Higher derivative corrections}

Now we turn to four-derivative terms.  It is known that (up to field redefinitions), the type II two-derivative supergravity action gets no corrections until certain eight-derivative terms predicted from string theory appear.  Thus the action receives only $(\a')^3$ corrections, and is uncorrected at order $\a'$ and $(\a')^2$.  It then follows, trivially, that the Buscher rules which we wrote down before continue to be symmetries of (the NS-NS part of) the action to order $(\a')^2$.

We will then assume that this observation holds also in the presence of branes, where suddenly the idea that the Buscher rules remain uncorrected at order $(\a')^2$ becomes a powerful tool.  The worldvolume actions of D-branes, and the Wess-Zumino piece in particular, is known to receive four-derivative corrections at order $(\a')^2$.  If the original Buscher rules continue to describe T-duality at this order, then they can be used to strongly constrain these corrections to the action, since the four-derivative parts of the action will need to be T-duality covariant by themselves.  On the other hand, if the Buscher rules {\it{were}} corrected to this order, then it would be much more difficult to extract any useful information, since we would have to contend with mixing between T-duality transformations of the zero-derivative and four-derivative parts of the action.

It's not completely clear that our assumption is reasonable - one could perhaps imagine corrections to the Buscher rules which were non-vanishing only in the presence of branes or other sources.  However, for now we will proceed with this idea, and we will find that the computations we do later in section \ref{sec:DiscAmplitudes} will confirm the predictions we make here, thus justifying, to some extent, our assumptions.

Now we turn to the known corrections to the Wess-Zumino action
\cite{Bershadsky:1995,Green:1996dd,Cheung:1997az,Minasian:1997mm},
\be \mathcal{L}_{WZ}=Ce^B\lp 1+\frac{\pi^2(\a')^2}{24}\lp\tr R_T\w
R_T-\tr R_N\w R_N\rp+\cdots\rp. \ee The order $(\a')^2$ correction
here is proportional to a four-form
\begin{multline}
\label{eq:OriginalX4}
X^{(4)}_{\mathrm{original}}=\tr R_T\w R_T-\tr R_N\w R_N \\ =\frac{1}{4}\lp-g_T^{eg}g_T^{fh}\lp R_T\rp_{abef}\lp R_T\rp_{cdgh}+\d^{ik}\d^{j\ell}\lp R_N\rp_{ab}^{\hphantom{ab}ij}\lp R_N\rp_{cd}^{\hphantom{cd}k\ell}\rp dx^a\w dx^b\w dx^c\w dx^d,
\end{multline}
where $g_T$ is the induced metric on the brane worldvolume, $R_T$ is the curvature tensor built from $g_T$, and $R_N$ is the curvature of the normal bundle.  Here and throughout this paper we use the indices $a$, $b$, etc. to refer to the worldvolume of the D-brane, and indices $i$, $j$, etc. to refer to the normal bundle.  Our notation largely follows that of~\cite{Bachas:1999um}.  We will use indices $\m$, $\n$, etc. for the ten-dimensional spacetime.  If the brane positions are given by $X^\m(x^a)$, then we have $(g_T)_{ab}=g_{\m\n}\p_aX^\m\p_bX^\n$, and we can pick an orthonormal frame $\xi^\m_i$ for the normal bundle which satisfies $g_{\m\n}\xi^\m_i\xi^\n_j=\d_{ij}$ and $g_{\m\n}\p_aX^\m\xi^\n_i=0$.

In order to relate the curvatures $R_T$ and $R_N$ to the ten-dimensional spacetime curvature, we must first introduce the second fundamental form~\cite{Eisenhart},
\be
\O_{ab}^i=\d^{ij}g_{\m\n}\xi^\m_j\lp\p_a\p_bX^\n-\lp\G_T\rp^c_{ab}\p_cX^\n+\G^\n_{\rho\s}\p_aX^\rho\p_bX^\s\rp.
\ee
In this expression, $\G^\n_{\rho\s}$ and $(\G_T)^c_{ab}$ are the Christoffel symbols constructed from the spacetime and worldvolume metrics respectively.

We then use the Gauss-Codazzi equations, which state
\beq
\lp R_T\rp_{abcd} &=& R_{abcd}+\d_{ij}\lp\O^i_{ac}\O^j_{bd}-\O^i_{ad}\O^j_{bc}\rp,\non\\
\lp R_N\rp_{ab}^{\hphantom{ab}ij} &=& -R_{ab}^{\hphantom{ab}ij}+g_T^{cd}\lp\O^i_{ac}\O^j_{bd}-\O^j_{ac}\O^i_{bd}\rp.
\eeq
Here we raise and lower indices with $(g_T)_{ab}$ or $\d_{ij}$, as appropriate, and we pull back indices from spacetime using either $\p_aX^\m$ or $\xi^\m_i$, so
\be
R_{abcd}=\pa_aX^\m\pa_bX^\n\pa_cX^\rho\pa_dX^\s R_{\m\n\rho\s},\qquad R_{ab}^{\hphantom{ab}ij}=\d^{ik}\d^{j\ell}\p_aX^\m\p_bX^\n\xi^\rho_k\xi^\s_\ell R_{\m\n\rho\s}.
\ee

We will work in a linearized approximation, which means that we expand all of our fields around a flat background and work to leading order in the fluctuations.  We do this both to greatly simplify our calculations, and also because these are really the only results that we can realistically compare to the disc amplitudes we compute in section \ref{sec:DiscAmplitudes}.  Fortunately, this does provide an enormous simplification since the second fundamental form vanishes in the flat background and so must be at least first order in fluctuations, which means that it contributes to $R_T$ and $R_N$ only at second order in the fields or higher.  Meanwhile, the spacetime curvature does have a piece which is first order in the fluctuations,
\be
R_{\m\n\rho\s}=-\p_{\m[\rho}h_{\s]\n}+\p_{\n[\rho}h_{\s]\m}+\mathcal{O}(h^2),
\ee
where we have split the metric into background plus fluctuation, $g_{\m\n}=\eta_{\m\n}+h_{\m\n}$.  Thus, to leading order in the fluctuations,
\begin{multline}
\label{eq:OriginalX4Linearized}
\lp X^{(4)}_{\mathrm{original}}\rp_{abcd} \\
=12\lp-\p_{[a}^{\hphantom{[a}e}h_b^{\hphantom{b}f}\p_{c|e|}h_{d]f}+\p_{[a}^{\hphantom{[a}e}h_b^{\hphantom{b}f}\p_{c|f|}h_{d]e}+\p_{[a}^{\hphantom{[a}i}h_b^{\hphantom{b}j}\p_{c|i|}h_{d]j}-\p_{[a}^{\hphantom{[a}i}h_b^{\hphantom{b}j}\p_{c|j|}h_{d]i}\rp+\mathcal{O}(h^3).
\end{multline}

\subsection{T-dualizing the corrections}

Now we note that the action so far (to this order in $\a'$) is not consistent with T-duality, since
\be
\mathcal{L}_{WZ}^{(p+1)}=\frac{\pi^2\lp\a'\rp^2}{24}\lp Ce^B\rp^{(p-3)}\w X^{(4)}_{\mathrm{original}}
\ee
does not satisfy (\ref{eq:WZTDuality}).  In order to find an action that is consistent with T-duality, we make the following ansatz\footnote{The normalizations here are chosen so as to make the T-duality rules in (\ref{eq:XTR2}) simple.  In principle we could also include terms with $X^{(1)\,i_1i_2i_3}_a$ and $X^{(0)\,i_1i_2i_3i_4}$, which would in turn correspond to couplings of higher degree forms $C^{(p+3)}$ and $C^{(p+5)}$ to the D-brane.  However, it turns out that these couplings do not occur in the T-duality invariants built from $X^{(4)}_{\mathrm{original}}$.}
\beq
\label{eq:L(p+1)}
\frac{24}{\pi^2\lp\a'\rp^2}\mathcal{L}^{(p+1)}_{a_1\cdots a_{p+1}} &=& \frac{\lp p+1\rp!}{4!\lp p-3\rp!}\lp Ce^B\rp^{(p-3)}_{[a_1\cdots a_{p-3}}X^{(4)}_{a_{p-2}a_{p-1}a_pa_{p+1}]}\non\\
&& +\frac{\lp p+1\rp!}{3!\lp p-2\rp!}\lp Ce^B\rp^{(p-1)}_{[a_1\cdots a_{p-2}|i|}X^{(3)\,i}_{a_{p-1}a_pa_{p+1}]}\\
&& +\frac{\lp p+1\rp!}{2^2\lp p-1\rp!}\lp Ce^B\rp^{(p+1)}_{[a_1\cdots a_{p-1}|i_1i_2|}X^{(2)\,i_1i_2}_{a_pa_{p+1}]}\non
\eeq
We assume that the objects $X^{(n)}$ are built out of NS-NS sector closed string fields\footnote{Note that the Buscher rules always preserve the number of R-R fields which appear in an expression, so this Wess-Zumino term does not mix under T-duality with terms that contain no R-R fields, such as DBI, or with terms that contain more than one R-R field.}.

To impose consistency under T-duality, we must ensure that this ansatz satisfies (\ref{eq:WZTDuality}), which happens iff
\be
\label{eq:XTR1}
X^{(4)\prime}_{a_1a_2a_3a_4}=X^{(4)}_{a_1a_2a_3a_4},\qquad X^{(3)\prime\,i}_{a_1a_2a_3}=X^{(3)\,i}_{a_1a_2a_3},\qquad X^{(2)\prime\,i_1i_2}_{a_1a_2}=X^{(2)\,i_1i_2}_{a_1a_2},
\ee
and\footnote{Here the T-duality transformation swaps an upper $y$ index with a lower $y$ index (though of course at linearized order around a flat background this is irrelevant).  This is a frequent feature of T-duality transformations of NS-NS fields and fluxes, such as for example so-called generalized NS-NS fluxes~\cite{Shelton:2005cf}.}
\be
\label{eq:XTR2}
X^{(3)\prime\,y}_{a_1a_2a_3}=X^{(4)}_{a_1a_2a_3y},\qquad X^{(2)\prime\,iy}_{a_1a_2}=X^{(3)\,i}_{a_1a_2y},
\ee
where a prime means that we have used the Buscher rules to transform the object in question.  This ansatz and these consistency conditions should in fact hold even beyond the linearized approximation, though at higher orders we may also have to incorporate open string fields.

Now we would like to build an action which includes the known terms (\ref{eq:OriginalX4}) but which is consistent with the T-duality rules expressed above.  Note that all four of the terms in (\ref{eq:OriginalX4Linearized}) have two of the four antisymmetrized free indices attached to derivatives.  The Buscher rules, given our assumption that they are exact to this order in $\a'$, will preserve this fact - any terms which can mix with these four terms under T-duality must also have two of the antisymmetrized indices occupied by derivatives.  One immediate consequence of this is that we need not consider terms in $X^{(n)}$ which are linear order in NS-NS fluctuations, since in that case all derivatives would be hitting the same field and antisymmetrizing any two derivatives would give zero.  This is not to say that terms with only one NS-NS field will not occur (indeed they are expected, see \cite{Garousi:2010ki}), but simply that they cannot appear in the same T-duality invariant as (\ref{eq:OriginalX4Linearized}).  Furthermore, applying the Buscher rules never reduces the number of fluctuations in a term, so we see that we can restrict ourselves to terms which are quadratic in the fluctuations and we can also restrict ourselves to the linearized version of the Buscher rules,
\be
h_{yy}'=-h_{yy},\qquad h_{\m y}'=B_{\m y},\qquad B_{\m y}'=h_{\m y},\qquad\Phi'=\Phi-\hlf h_{yy},
\ee
with $h_{\m\n}$ and $B_{\m\n}$ left invariant.

Under these transformations, it is not hard to verify that the terms in (\ref{eq:OriginalX4Linearized}) can only mix with certain terms, which we can enumerate,
\beq
X^{(4)}_{a_1a_2a_3a_4} &=& \a_1\p_{[a_1}^{\hphantom{[a_1}b}h_{a_2}^{\hphantom{a_2}c}\p_{a_3|b|}h_{a_4]c}+\a_2\p_{[a_1}^{\hphantom{[a_1}b}h_{a_2}^{\hphantom{a_2}c}\p_{a_3|c|}h_{a_4]b}+\a_3\p_{[a_1}^{\hphantom{[a_1}j}h_{a_2}^{\hphantom{a_2}k}\p_{a_3|j|}h_{a_4]k} \non\\
&& +\a_4\p_{[a_1}^{\hphantom{[a_1}j}h_{a_2}^{\hphantom{a_2}k}\p_{a_3|k|}h_{a_4]j}+\a_5\p_{[a_1}^{\hphantom{[a_1}b}B_{a_2}^{\hphantom{a_2}j}\p_{a_3|b|}B_{a_4]j}+\a_6\p_{[a_1}^{\hphantom{[a_1}j}B_{a_2}^{\hphantom{a_2}b}\p_{a_3|j|}B_{a_4]b},\non\\
X^{(3)\,i}_{a_1a_2a_3} &=& \b_1\p_{[a_1}^{\hphantom{[a_1}b}h_{a_2}^{\hphantom{a_2}c}\p_{a_3]b}B^i_{\hphantom{i}c}+\b_2\p_{[a_1}^{\hphantom{[a_1}b}h_{a_2}^{\hphantom{a_2}c}\p_{a_3]c}B^i_{\hphantom{i}b}+\b_3\p_{[a_1}^{\hphantom{[a_1}j}h_{a_2}^{\hphantom{a_2}k}\p_{a_3]j}B^i_{\hphantom{i}k} \non\\
&& +\b_4\p_{[a_1}^{\hphantom{[a_1}j}h_{a_2}^{\hphantom{a_2}k}\p_{a_3]k}B^i_{\hphantom{i}j}+\b_5\p_{[a_1}^{\hphantom{[a_1}b}h^{ij}\p_{a_2|b|}B_{a_3]j}+\b_6\p_{[a_1}^{\hphantom{[a_1}j}h^{ib}\p_{a_2|j|}B_{a_3]b},\non\\
X^{(2)\,i_1i_2}_{a_1a_2} &=& \g_1\p_{[a_1}^{\hphantom{[a_1}b}h^{[i_1|j|}\p_{a_2]b}h^{i_2]}_{\hphantom{i_2]}j}+\g_2\p_{[a_1}^{\hphantom{[a_1}j}h^{[i_1|b|}\p_{a_2]j}h^{i_2]}_{\hphantom{i_2]}b}+\g_3\p_{[a_1}^{\hphantom{[a_1}b}B^{[i_1|c|}\p_{a_2]b}B^{i_2]}_{\hphantom{i_2]}c} \non\\
&& +\g_4\p_{[a_1}^{\hphantom{[a_1}b}B^{[i_1|c|}\p_{a_2]c}B^{i_2]}_{\hphantom{i_2]}b}+\g_5\p_{[a_1}^{\hphantom{[a_1}j}B^{[i_1|k|}\p_{a_2]j}B^{i_2]}_{\hphantom{i_2]}k}+\g_6\p_{[a_1}^{\hphantom{[a_1}j}B^{[i_1|k|}\p_{a_2]k}B^{i_2]}_{\hphantom{i_2]}j},\non
\eeq

From (\ref{eq:OriginalX4Linearized}) we know that $-\a_1=\a_2=\a_3=-\a_4=12$, but we would like to use our T-duality constraints to determine the remaining fourteen constants.  To proceed, we need to evaluate the expressions above in an ansatz with a circle bundle.  For instance, suppose the circle bundle is along the brane, then we would evaluate $X^{(4)}$ as
\be
X^{(4)}_{a_1a_2a_3a_4}=\widehat{X}^{(4)}_{a_1a_2a_3a_4}+\a_1\p_{[a_1}^{\hphantom{[a_1}\hat{b}}h_{a_2|y|}\p_{a_3|\hat{b}|}h_{a_4]y}+\a_6\p_{[a_1}^{\hphantom{[a_1}j}B_{a_2|y|}\p_{a_3|j|}B_{a_4]y},
\ee
where hatted indices are summed over all directions along the brane excluding $y$, and where $\widehat{X}^{(4)}$ represents the expression for $X^{(4)}$ but with $y$ excluded from all sums.  Under T-duality, this expression becomes
\be
\label{eq:X4pT}
X^{(4)\,\prime}_{a_1a_2a_3a_4}=\widehat{X}^{(4)}_{a_1a_2a_3a_4}+\a_1\p_{[a_1}^{\hphantom{[a_1}b}B_{a_2|y|}\p_{a_3|b|}B_{a_4]y}+\a_6\p_{[a_1}^{\hphantom{[a_1}\hat{\jmath}}h_{a_2|y|}\p_{a_3|\hat{\jmath}|}h_{a_4]y}.
\ee
Meanwhile, if the circle bundle is normal to the brane we have
\be
\label{eq:X4n}
X^{(4)}_{a_1a_2a_3a_4}=\widehat{X}^{(4)}_{a_1a_2a_3a_4}+\a_3\p_{[a_1}^{\hphantom{[a_1}\hat{\jmath}}h_{a_2|y|}\p_{a_3|\hat{\jmath}|}h_{a_4]y}+\a_5\p_{[a_1}^{\hphantom{[a_1}b}B_{a_2|y|}\p_{a_3|b|}B_{a_4]y}.
\ee
Comparing (\ref{eq:X4pT}) and (\ref{eq:X4n}) we learn that $\a_1=\a_5$ and $\a_6=\a_3$.  Similar considerations for $X^{(3)}$ and $X^{(2)}$ show that $\b_1=\b_5$, $\b_6=\b_3$, $\g_2=\g_5$, and $\g_3=\g_1$.

Next, we also compute
\beq
X^{(4)\,\prime}_{a_1a_2a_3y} &=& \hlf \a_1\lp\p_{[a_1}^{\hphantom{[a_1}b}h_{a_2}^{\hphantom{a_2}c}\p_{a_3]b}B_{cy}-\p_{[a_1}^{\hphantom{[a_1}b}h_{|yy|}\p_{a_2|b|}B_{a_3]y}\rp+\hlf \a_2\p_{[a_1}^{\hphantom{[a_1}b}h_{a_2}^{\hphantom{a_2}c}\p_{a_3]c}B_{by} \non\\
&& \qquad+\hlf \a_3\p_{[a_1}^{\hphantom{[a_1}\hat{j}}h_{a_2}^{\hphantom{a_2}\hat{k}}\p_{a_3]\hat{j}}B_{\hat{k}y}+\hlf \a_4\p_{[a_1}^{\hphantom{[a_1}\hat{j}}h_{a_2}^{\hphantom{a_2}\hat{k}}\p_{a_3]\hat{k}}B_{\hat{j}y} \non\\
&& \qquad-\hlf \a_5\p_{[a_1}^{\hphantom{[a_1}b}h^{\hat{j}}_{\hphantom{\hat{j}}y}\p_{a_2|b|}B_{a_3]\hat{j}}-\hlf \a_6\p_{[a_1}^{\hphantom{[a_1}\hat{j}}h^b_{\hphantom{b}y}\p_{a_2|\hat{j}|}B_{a_3]b},
\eeq
and
\beq
X^{(3)\,y}_{a_1a_2a_3} &=& -\b_1\p_{[a_1}^{\hphantom{[a_1}b}h_{a_2}^{\hphantom{a_2}c}\p_{a_3]b}B_{cy}-\b_2\p_{[a_1}^{\hphantom{[a_1}b}h_{a_2}^{\hphantom{a_2}c}\p_{a_3]c}B_{by}-\b_3\p_{[a_1}^{\hphantom{[a_1}\hat{\jmath}}h_{a_2}^{\hphantom{a_2}\hat{k}}\p_{a_3]\hat{\jmath}}B_{\hat{k}y} \non\\
&& \qquad-\b_4\p_{[a_1}^{\hphantom{[a_1}\hat{\jmath}}h_{a_2}^{\hphantom{a_2}\hat{k}}\p_{a_3]\hat{k}}B_{\hat{\jmath}y}+\b_5\lp\p_{[a_1}^{\hphantom{[a_1}b}h^{\hat{\jmath}}_{\hphantom{\hat{\jmath}}|y|}\p_{a_2|b|}B_{a_3]\hat{j}}+\p_{[a_1}^{\hphantom{[a_1}b}h_{|yy|}\p_{a_2|b|}B_{a_3]y}\rp \non\\
&& \qquad+\b_6\p_{[a_1}^{\hphantom{[a_1}\hat{\jmath}}h^b_{\hphantom{b}|y|}\p_{a_2|\hat{\jmath}|}B_{a_3]b},
\eeq
from which we deduce that $\b_1=-\hlf \a_1$, $\b_2=-\hlf \a_2$, $\b_3=-\hlf \a_3$, $\b_4=-\hlf \a_4$, $\b_5=-\hlf \a_5=-\hlf \a_1$, and $\b_6=-\hlf \a_6=-\hlf \a_3$.

A comparison of $X^{(3)\,i\,\prime}_{a_1a_2y}$ and $X^{(2)\,iy}_{a_1a_2}$ then lead us also to $\g_1=-\frac{1}{3}\b_5=-\frac{1}{3}\b_1$, $\g_2=-\frac{1}{3}\b_6$, $\g_3=-\frac{1}{3}\b_1$, $\g_4=-\frac{1}{3}\b_2$, $\g_5=-\frac{1}{3}\b_3$, and $\g_6=-\frac{1}{3}\b_4$.  Note that all the conditions are self-consistent, and we are left with the result,
\beq
X^{(4)}_{a_1a_2a_3a_4} &=& 12\lp -\p_{[a_1}^{\hphantom{[a_1}b}h_{a_2}^{\hphantom{a_2}c}\p_{a_3|b|}h_{a_4]c}+\p_{[a_1}^{\hphantom{[a_1}b}h_{a_2}^{\hphantom{a_2}c}\p_{a_3|c|}h_{a_4]b}+\p_{[a_1}^{\hphantom{[a_1}j}h_{a_2}^{\hphantom{a_2}k}\p_{a_3|j|}h_{a_4]k}\right. \non\\
&& \left.-\p_{[a_1}^{\hphantom{[a_1}j}h_{a_2}^{\hphantom{a_2}k}\p_{a_3|k|}h_{a_4]j}-\p_{[a_1}^{\hphantom{[a_1}b}B_{a_2}^{\hphantom{a_2}j}\p_{a_3|b|}B_{a_4]j}+\p_{[a_1}^{\hphantom{[a_1}j}B_{a_2}^{\hphantom{a_2}b}\p_{a_3|j|}B_{a_4]b}\rp,\non\\
X^{(3)\,i}_{a_1a_2a_3} &=& 6\lp\p_{[a_1}^{\hphantom{[a_1}b}h_{a_2}^{\hphantom{a_2}c}\p_{a_3]b}B^i_{\hphantom{i}c}-\p_{[a_1}^{\hphantom{[a_1}b}h_{a_2}^{\hphantom{a_2}c}\p_{a_3]c}B^i_{\hphantom{i}b}-\p_{[a_1}^{\hphantom{[a_1}j}h_{a_2}^{\hphantom{a_2}k}\p_{a_3]j}B^i_{\hphantom{i}k}\right. \non\\
&& \left.+\p_{[a_1}^{\hphantom{[a_1}j}h_{a_2}^{\hphantom{a_2}k}\p_{a_3]k}B^i_{\hphantom{i}j}+\p_{[a_1}^{\hphantom{[a_1}b}h^{ij}\p_{a_2|b|}B_{a_3]j}-\p_{[a_1}^{\hphantom{[a_1}j}h^{ib}\p_{a_2|j|}B_{a_3]b}\rp,\\
X^{(2)\,i_1i_2}_{a_1a_2} &=& 2\lp -\p_{[a_1}^{\hphantom{[a_1}b}h^{[i_1|j|}\p_{a_2]b}h^{i_2]}_{\hphantom{i_2]}j}+\p_{[a_1}^{\hphantom{[a_1}j}h^{[i_1|b|}\p_{a_2]j}h^{i_2]}_{\hphantom{i_2]}b}-\p_{[a_1}^{\hphantom{[a_1}b}B^{[i_1|c|}\p_{a_2]b}B^{i_2]}_{\hphantom{i_2]}c}\right. \non\\
&& \left.+\p_{[a_1}^{\hphantom{[a_1}b}B^{[i_1|c|}\p_{a_2]c}B^{i_2]}_{\hphantom{i_2]}b}+\p_{[a_1}^{\hphantom{[a_1}j}B^{[i_1|k|}\p_{a_2]j}B^{i_2]}_{\hphantom{i_2]}k}-\p_{[a_1}^{\hphantom{[a_1}j}B^{[i_1|k|}\p_{a_2]k}B^{i_2]}_{\hphantom{i_2]}j}\rp,\non
\eeq
Taking into account the factorial factors in (\ref{eq:L(p+1)}), we see that this result can be written in the form (\ref{eq:FixedCouplings}).

\subsection{Brief discussion}

What we have argued in this section is that considerations of
T-duality combined with the previously known four-derivative
contributions to the Wess-Zumino action are sufficient to fix the
coefficients of the eighteen couplings listed above.  This is
certainly not to claim that these will be the only four-derivative
corrections to this action.  Indeed, the action as we would write it
now, though consistent with linearized T-duality, would not be
consistent with R-R gauge invariance, $B$-field gauge invariance, or
diffeomorphism invariance, even at the linearized level, nor would
it be completely consistent with T-duality at the non-linear level.
In order to restore consistency, then, many more terms will have to
be included.  Moreover, it is certainly desirable to rewrite the
new couplings to display their topological nature if possible.
Additional four derivative terms, beyond those we are presenting are
required. These can be determined by computing further string disc
amplitudes. A vertex operator computation, however, can rapidly
become very cumbersome to say the least since higher-point
correlation functions will be needed. A possible approach could be
to rederive the new couplings using an anomaly inflow argument along the lines of~\cite{Green:1996dd}.
Work in this direction is in progress~\cite{inprogress}. Nonetheless, the presence of
these extra terms cannot modify the coefficients that we have
determined above.

However, our derivation is perhaps not yet completely satisfactory.
We have relied crucially on the assumption that the Buscher rules do
not get modified at this order in the derivative expansion.  So, we
would like to check our results by comparing them with actual
computations of string disc amplitudes, which will be our task in
the remainder of the paper.  The relevant disc amplitudes, involving
the insertion of one R-R vertex operator and two NS-NS vertex
operators, are subtle objects, and in order to accurately compute
the full amplitudes one must deal with issues involving spurious
states not decoupling. Indeed, as discussed in~\cite{Green:1987qu},
the OPE of two physical vertex operators can lead to vertex
operators of unphysical states. The coefficients of these states
are, however, total derivatives which decouple on the sphere. On
spaces like the disc these can lead to boundary terms. If a
particular term receives contributions from diagrams in which these
spurious states propagate, then the naive computation is unreliable,
in the sense that amplitudes computed in different pictures (i.e.
with different choices of how to distribute picture charge) can
disagree. Moreover, poles in the small momentum expansion corresponding to spurious states may appear.
Such poles appear, for example, in the two-point function of one R-R potential and one $B$-field, which
in the supergravity limit contributes to the $\int C^{(p-1)} \wedge B$ coupling on a D$p$-brane. It is
not difficult to see that neither of the two diagrams contributing (one with a vertex directly on the
brane and one with a vertex in spacetime) can have any poles in the momentum expansion. However,
depending on the picture being used to compute the result poles at small momentum can appear. The presence of
these poles has first been noticed in~\cite{Craps:1998fn,Craps:2000zr}.
Presumably the discrepancies can be repaired by adding appropriate
boundary terms. Work to verify this is in progress.
\begin{figure}[!htb]
\label{fig:Diagrams}
\begin{center}
\scalebox{.65}{\includegraphics{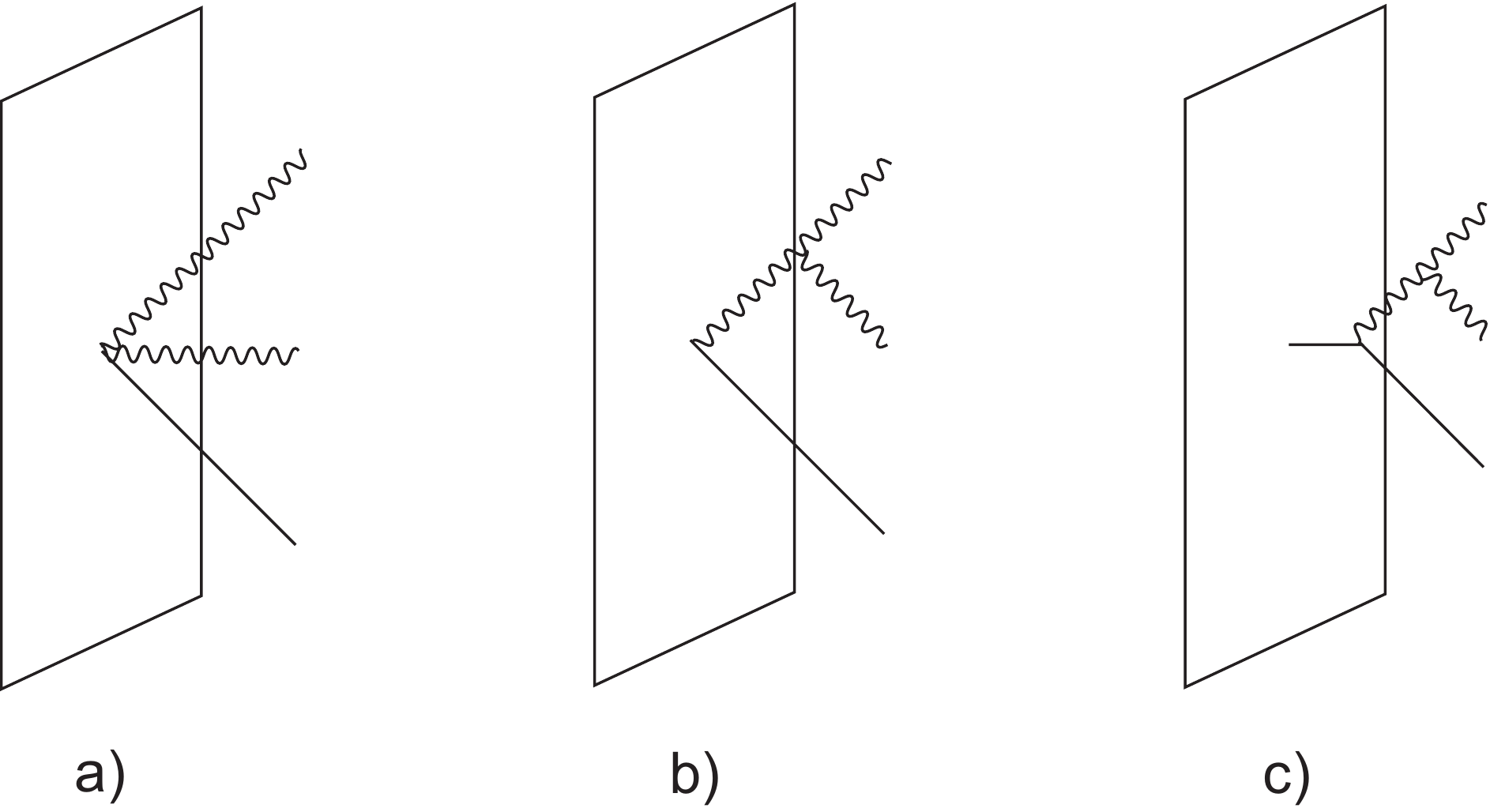}} \caption{In these
diagrams the wavy lines represent NS-NS fields while the solid lines
represent R-R fields.}
\end{center}
\end{figure}

Fortunately, for the specific couplings that appear in
(\ref{eq:FixedCouplings}) these issues will not arise. Roughly, the
reason is that these couplings should only get contributions from
contact diagrams in which all three fields meet in a vertex on the
brane as in figure 1a), and not from diagrams in which states propagate between two
vertices, such as the examples shown in figures 1b) and 1c).
If there were contributions of the latter type, then
spurious states propagating between the vertices could potentially
cause problems, but the former type of coupling should be safe. For
example, consider the terms in (\ref{eq:OriginalX4Linearized}).
Since bulk spacetime interactions between gravitons and R-R fields
cannot change the degree of the R-R field, and since we know that
the bulk vertices and propagators do not get four-derivative
corrections, the only possible diagrams with internal bulk
proagators that can contribute are those in which a $(p-3)$-form R-R
field has a four-derivative contact term on the brane with either
zero or one graviton, like in figure 1c) or 1b) respectively.  But it is easy to see that we can't write
down any such non-vanishing contact terms, since we would need
momenta to soak up three of the indices along the brane (we could
soak up at most $p-3$ indices with the R-R form and one index with
the graviton polarization), but there are only at most two
independent momenta available at the vertex.

Similar arguments can be used to show that the other terms in
(\ref{eq:FixedCouplings}) do not receive contributions from diagrams
with internal bulk propagators, but the details quickly become quite
involved.  Instead, let us proceed to compute the relevant couplings
in various pictures.  Since the result will turn out to be
independent of the picture, and will agree with our predictions from
T-duality, it is reasonable to conclude that spurious states are not
making troublesome contributions.

\section{Disc Amplitudes}
\label{sec:DiscAmplitudes}

In this section we shall compute disc amplitudes with one insertion of a R-R field and two insertions of NS-NS fields.  For the specific couplings in (\ref{eq:FixedCouplings}), we will perform the computation in various different pictures and we will confirm that they agree with each other and with T-duality.  We will also point out how the situation is more subtle for some other couplings, with unphysical poles arising in the amplitude, and picture-dependent results.

For the case of the original couplings (\ref{eq:OriginalWZAction}), the three-point amplitudes were computed in~\cite{Craps:1998fn,Stefanski:1998yx,Morales:1998ux}.

\subsection{Basic Conventions}

We work in the RNS worldsheet formalism, with the closed string vertex operators being constructed out of bosons $X^\m(z,\zbar)=X^\m(z)+\widetilde{X}^\m(\zbar)$ and fermions $\psi^\m(z)$, $\widetilde{\psi}^\m(\zbar)$, as well as the picture ghosts $\phi(z)$ and $\widetilde{\phi}(\zbar)$.  Since we work with integrated vertex operators, we won't need the $b$, $c$, $\eta$, $\xi$ ghosts.

On the upper half-plane, the holomorphic fields have OPEs among themselves\footnote{In this section we will mostly work in units where $\a'=2$, in order to keep things notationally simple.}
\begin{equation}
\label{eq:HolomorphicOPEs}
\begin{split}
X^\m(z) X^\n(w) & \sim - \eta^{\m\n} \log(z-w), \cr
\psi^\m(z) \psi^\n(w) & \sim - {\eta^{\m\n}\over  z-w}, \cr
\phi(z) \phi(w) & \sim - \log (z-w),\cr
\end{split}
\end{equation}
with similar expressions for the antiholomorphic fields.  The presence of the boundary (the real axis), representing the D-brane, leads to non-trivial OPEs also between holomorphic and antiholomorphic fields,
\begin{equation}
\begin{split}
X^\m(z) \widetilde{X}^\n(\bar{w}) & \sim - D^{\m\n} \log(z-\bar{w}), \cr
\psi^\m(z) \widetilde{\psi}^\n(\bar{w}) & \sim - {D^{\m\n}\over  z-\bar{w}}, \cr
\phi(z) \widetilde{\phi}(\bar{w}) & \sim - \log (z-\bar{w}).\cr
\end{split}
\end{equation}
Here the matrix $D^{\m\n}$ is a diagonal matrix that agrees with $\eta^{\m\n}$ in directions along the brane (Neumann boundary conditions) and with $-\eta^{\m\n}$ in directions normal to the brane (Dirichlet boundary conditions).  In our previous notation, $D^{ab}=\eta^{ab}$, $D^{ij}=-\d^{ij}$, $D^{ai}=0$.  Using $\eta_{\m\n}$ to raise or lower indices, then we have $D^\m_{\hphantom{\m}\rho}D^\rho_{\hphantom{\rho}\n}=\d^\m_\n$.  One can now use a convenient trick~\cite{Garousi:1996ad,Gubser:1996wt} when computing amplitudes.  One can make the replacements
\be
\label{eq:ReplacementTrick}
\widetilde{X}^\m(\zbar)\rightarrow D^\m_{\hphantom{\m}\n}X^\n(\zbar),\qquad\widetilde{\psi}(\zbar)\rightarrow D^\m_{\hphantom{\m}\n}\psi^\n(\zbar),\qquad\widetilde{\phi}(\zbar)\rightarrow\phi(\zbar),
\ee
and then use only the holomorphic OPEs (\ref{eq:HolomorphicOPEs}), but where we now regard $z$ and $\zbar$ as independent insertion points.

In order to construct R-R vertex operators, we will also need spin fields $S_A(z)$ and $\widetilde{S}_B(\zbar)$, where $A$ and $B$ are spinor indices.  Rather than give the individual OPEs involving spin fields, it will suffice to quote the general fermion sector expectation values that we will need\footnote{A similar expression appears in~\cite{Liu:2001qa}, though their result restricts to fermions on the boundary of the disc.  We need the more general result shown here.},
\begin{multline}
\left<S_A(z) \widetilde{S}_B(\bar z)\psi^{\m_1} (z_1) \dots \psi^{\m_n}(z_n)  \right> =
 {1\over 2^{n/2}} {(z-\bar z)^{n/2-5/4}\over \sqrt{(z_1 - z )(z_1 - \bar z) \dots (z_n - z)(z_n - \bar z)} } \\
\times\Big[ (\Gamma^{\m_n \dots \m_1} {\cal C}^{-1}M^T)_{AB}+ \widehat{\psi^{\m_1} (z_1) \psi^{\m_2}(z_2)}(\Gamma^{\m_n \dots \m_3} {\cal C}^{-1}M^T)_{AB} \pm
\dots \\
+ \widehat{\psi^{\m_1}(z_1) \psi^{\m_2} (z_2) }
\widehat{\psi^{\m_3}(z_3) \psi^{\m_4} (z_4) } (\Gamma^{\m_n \dots \m_5} {\cal C}^{-1}M^T)_{AB} \pm
\dots \Big],
\end{multline}
where
\begin{equation}
\widehat{\psi^{\m_i}(z_i) \psi^{\m_j} (z_j) }=-\eta^{\m_i \m_j}
{(z_i-z)(z_j-\bar z)+(z_j - z)(z_i-\bar z)\over (z_i-z_j)(z-\bar z)}=-\frac{\eta^{\m_i\m_j}}{z-\zbar}\mathcal{P}(z_i,z_j).
\end{equation}
In these expressions we use real symmetric $32\times 32$ gamma matrices $(\G^\m)_A^{\hphantom{A}B}$ which satisfy
\be
\left\{\G^\m,\G^\n\right\}=-2\eta^{\m\n},
\ee
$\mathcal{C}^{AB}$ is an antisymmetric charge conjugation matrix, and $M_A^{\hphantom{A}B}$ encodes the Neumann and Dirichlet boundary conditions as they are realized on spinor indices, so that it satisfies $\G^\m M=D^\m_{\hphantom{\m}\n}M\G^\n$.  It is explicitly given by
\be
M=\left\{\begin{array}{ll}\pm\frac{i}{(p+1)!}\lp\e^v\rp_{a_0\cdots a_p}\G^{a_0}\cdots\G^{a_p}, & \mathrm{for\ }p\mathrm{\ even}, \\ \pm\frac{1}{(p+1)!}\lp\e^v\rp_{a_0\cdots a_p}\G^{a_0}\cdots\G^{a_p}\G_{11}, & \mathrm{for\ }p\mathrm{\ odd},\end{array}\right.
\ee
where $\e^v$ is the epsilon tensor on the brane worldvolume and where
\be
\G_{11}=\frac{1}{10!}\e_{\m_0\cdots\m_9}\G^{\m_0}\cdots\G^{\m_9}=\G^0\cdots\G^9.
\ee
We will not be attempting to compute the overall normalization of our result (as opposed to relative phases, which will of course be crucial), so we can freely ignore the $\pm 1$ or $\pm i$ in the definition of $M$.

One final useful fact is that
\be
\label{eq:GammaTranspose}
\mathcal{C}^{-1}\lp\G^\m\rp^T\mathcal{C}=-\G^\m.
\ee

Though it is convenient to evaluate the expectation values in the upper half-plane, parametrized by $z$, it turns out that to perform the integrals over the vertex operator positions it is more convenient to map the results to the interior of the unit disc with coordinate $w$, via the map
\be
w=\frac{i-z}{i+z},\qquad z=i\frac{1-w}{1+w}.
\ee
We will also use the $\operatorname{SL}(2,\R)$ invariance to fix $w_0=0$, and $w_1=r_1$, leaving $w_2=r_2e^{i\th}$.  Here $0\le r_i\le 1$ and $0\le\th\le 2\pi$.  Taking into account the measure factors $d^2z$ and the Jacobian factor from fixing $\operatorname{SL}(2,\R)$, and mapping to the disc, we are left with the measure
\be
\sim\int_0^{2\pi} d\th\int_0^1\frac{r_1dr_1}{\lp 1+r_1\rp^4}\int_0^1\frac{r_2dr_2}{\left| 1+r_2e^{i\th}\right|^4}.
\ee
As mentioned above, we will not concern ourselves with the overall normalization of the result, since at the end of the calculation this can be fixed by comparing to the known terms (\ref{eq:OriginalX4}).  The relative factors are the crucial pieces of new information that we would like to compare to the T-duality predictions (\ref{eq:FixedCouplings}).

Thus the amplitudes we will be computing are
\be
\mathcal{A}\sim\int_0^{2\pi} d\th\int_0^1\frac{r_1dr_1}{\lp 1+r_1\rp^4}\int_0^1\frac{r_2dr_2}{\left| 1+r_2e^{i\th}\right|^4}\left\langle V^{R-R}V_1^{NS-NS}V_2^{NS-NS}\right\rangle.
\ee
Here we evaluate the expectation value on the upper half-plane, map to the disc, and fix positions as described.

\subsection{Selection rules}
\label{subsec:SelectionRules}

We would like to compare the couplings of (\ref{eq:FixedCouplings}) with the amplitudes we compute.  Let us denote the two NS-NS polarizations by $\lp\e_1\rp_{\m\n}$ and $\lp\e_2\rp_{\m\n}$, where a graviton is represented by a symmetric traceless polarization $\e_{\m\n}=h_{\m\n}$, and a $B$-field is represented by an antisymmetric polarization $\e_{\m\n}=B_{\m\n}$.  Let $p_1$ and $p_2$ be the momenta associated to these two fields.

We can split the couplings into two classes, which we will denote class A and class B.  Class A couplings involve one contraction between the two NS-NS polarizations and one contraction between $p_1$ and $p_2$, while class B couplings involve each polarization contracting with the other field's momentum.  Note that in each of $X^{(4)}$, $X^{(3)}$, and $X^{(2)}$ we have four couplings of class A and two of class B.  Moreover, every term has one additional factor each of $p_1$ and $p_2$ carrying a free index along the brane.  Schematically,
\begin{list}{$\bullet$}{}
\item Class A: $\lp p_1\cdot p_2\rp\lp\e_1\cdot\e_2\rp_{\m\n}p_{1\,a}p_{2\,b}$,
\item Class B: $\lp\e_1\cdot p_2\rp_\m\lp\e_2\cdot p_1\rp_\n p_{1\,a}p_{2\,b}$.
\end{list}
Here the indices $\m$ and $\n$ would lie along the brane in $X^{(4)}$, normal to the brane in $X^{(2)}$, and one each in $X^{(3)}$.

In fact, the particular couplings of (\ref{eq:FixedCouplings}) obey more specific selection rules.  We will see that once we apply the trick (\ref{eq:ReplacementTrick}), the polarizations appear in the vertex operators in the combination $\lp\e D\rp_{\m\n}$ and contractions are only done with $\eta^{\m\n}$.  Then there are four different possibilities for the contraction $\lp\e_1\cdot\e_2\rp$ which appears in class A, depending on which pair of indices we contract,
\be
\lp D\e_1^T\e_2D\rp_{\m\n},\qquad\lp D\e_1^TD\e_2^T\rp_{\m\n},\qquad\lp\e_1D\e_2D\rp_{\m\n},\qquad\lp\e_1\e_2^T\rp_{\m\n},
\ee
where $\e^T$ is simply the transpose of the polarization $\e$.
Now observe that in $X^{(4)}$ and $X^{(2)}$ we have either two gravitons or two $B$-fields, while in $X^{(3)}$ we always have one of each.  Because of this one can check that all twelve class A couplings satisfy the selection rules
\be
\label{eq:ClassASelectionRules}
\lp D\e_1^T\e_2D\rp_{\m\n}=\lp\e_1\e_2^T\rp_{\m\n},\qquad\lp D\e_1^TD\e_2^T\rp_{\m\n}=\lp\e_1D\e_2D\rp_{\m\n}.
\ee

For the six terms in class B, we similarly note that the free index on a graviton is always along the brane, while the free index on a $B$-field is always normal to the brane, which leads to the selection rules
\be
\label{eq:ClassBSelectionRules}
\lp\e Dp\rp_\m=\lp pD\e D\rp_\m,\qquad\lp\e p\rp_\m=\lp p\e D\rp_\m.
\ee

We will use these rules, together with $p_{i\,a}=\lp Dp_i\rp_a$, in the disc amplitudes to compute the specific coefficients which appear in (\ref{eq:FixedCouplings}).

\subsection{Computation in the $(-3/2,-1/2)\times(0,0)\times(0,0)$ picture}
\label{subsec:AsymmetricPicture}

We will now perform the computation for a particular distribution of picture charge, in which both NS-NS vertex operators are in the $(0,0)$ picture while the R-R vertex operator is in the $(-3/2,-1/2)$ picture.

\subsubsection{Vertex Operators}

Explicitly, we have
\be
V^{(0,0)}_i(z_i,\zbar_i)=\lp\e_iD\rp_{\m\n}\lp\pa X^\m+i\lp p_i\psi\rp\psi^\m\rp e^{ip_iX}(z_i)\lp\pa X^\n+i\lp p_iD\psi\rp\psi^\n\rp e^{ip_iDX}(\zbar_i),
\ee
and~\cite{Billo:1998vr,Liu:2001qa}
\be
V^{(-3/2,-1/2)}(z_0,\zbar_0)=\lp\mathcal{C}P_-\slashed{C}\rp^{AB}e^{-\frac{3}{2}\phi(z_0)}e^{-\hlf\phi(\zbar_0)}S_A(z_0)\widetilde{S}_B(\zbar_0)e^{ip_0X(z_0)+ip_0DX(\zbar_0)},
\ee
where the NS-NS vertex operators include polarizations $(\e_i)_{\m\n}$ and momenta $p_i$, $i=1,2$, and where the R-R vertex operator has momentum $p_0$ and antisymmetric polarization $C^{(n+1)}_{\m_0\cdots\m_n}$.  We also make use of definitions
\be
P_\pm=\hlf\lp 1\pm\G_{11}\rp,\qquad\mathrm{and}\qquad\slashed{C}=\frac{1}{(n+1)!}C^{(n+1)}_{\m_0\cdots\m_n}\G^{\m_0\cdots\m_n}.
\ee
Note the identity
\be
\lp\mathcal{C}P_-\slashed{C}\rp^{AB}\lp\G^{\m_k\cdots\m_1}\mathcal{C}^{-1}M^T\rp_{AB}=\lp -1\rp^{k+1}\Tr\lp P_-\slashed{C}M\G^{\m_1\cdots\m_k}\rp.
\ee
We also have the facts
\beq
\label{eq:TraceEvaluations}
\Tr\lp P_-\slashed{C}M\G^{abcd}\rp &\sim& \frac{16}{\lp p-3\rp!}\lp\e^v\rp^{e_0\cdots e_{p-4}abcd}C^{(p-3)}_{e_0\cdots e_{p-4}},\non\\
\Tr\lp P_-\slashed{C}M\G^{abci}\rp &\sim& \frac{16}{\lp p-2\rp!}\lp\e^v\rp^{e_0\cdots e_{p-3}abc}C^{\rlap{$\scriptstyle{(p-1)}$}\hphantom{e_0\cdots e_{p-3}}i}_{e_0\cdots e_{p-3}},\\
\Tr\lp P_-\slashed{C}M\G^{abij}\rp &\sim& \frac{16}{\lp p-1\rp!}\lp\e^v\rp^{e_0\cdots e_{p-2}ab}C^{\rlap{$\scriptstyle{(p+1)}$}\hphantom{e_0\cdots e_{p-2}}ij}_{e_0\cdots e_{p-2}},\non
\eeq
where we use $\sim$ because we ignore an overall $p$-dependent phase\footnote{We are actually glossing over another subtlety here; because of the presence of $\G_{11}$ in the definition of $P_-$ (and possibly of $M$), there will be a second contribution in the trace which is in some sense the Poincare dual of the first.  For instance in the first line there could be an additional coupling which is
\be
\frac{16}{\lp 9-p\rp !}\lp\e^n\rp_{i_1\cdots i_{9-p}}C^{(13-p)\,i_1\cdots i_{9-p}abcd}.
\ee
We will ignore these extra couplings, since they are not of the form that we are looking for.}.  These will be necessary to compare our disc amplitude computations with the T-duality predictions of section \ref{sec:Predictions}.

Moreover, there are also certain physical state conditions which must be satisfied, namely that
\be
\label{eq:PhysicalStateConditions}
p_0^2=p_1^2=p_2^2=0,\qquad p_i^\m\lp\e_i\rp_{\m\n}=\lp\e_i\rp_{\m\n}p_i^\n=0,\qquad p_0^\m C_{\m\n_1\cdots\n_n}=0,
\ee
and we have momentum conservation along the brane, $p_0^a+p_1^a+p_2^a=0$, or equivalently
\be
\label{eq:MomentumConservation}
p_0+Dp_0+p_1+Dp_1+p_2+Dp_2=0,
\ee
(this is actually enforced by the expectation value of the exponentials in the different vertex operators).

Now we turn to the computation of the couplings in (\ref{eq:FixedCouplings}), first those in class A and then those in class B.  Note that in all cases we have exactly four free indices, either transverse or normal to the brane, which implies that we need to isolate contributions from the fermionic sector expectation value which contain precisely four gamma matrices.

\subsubsection{Class A couplings}

Since we need terms with four gamma matrices, we must focus on the terms in $\langle V^{(-3/2,-1/2)}V_1^{(0,0)}V_2^{(0,0)}\rangle$ which have at least four fermions.  For the couplings in which $\e_1$ and $\e_2$ contract with each other, it turns out that contributions can come only from certain contractions of the eight fermion term, and from some of the four fermion terms in which the bosons, one from each NS-NS operator, contract with each other.

From the eight fermion term, it turns out that we have a factorization between the choice of polarization contraction and the choice of momentum contraction, so that all together the expectation value is
\begin{multline}
-\frac{\mathcal{K}}{16}\left|z_0-z_1\right|^{-2}\left|z_0-\zbar_1\right|^{-2}\left|z_0-z_2\right|^{-2}\left|z_0-\zbar_2\right|^{-2}\Tr\lp P_-\slashed{C}M\G^{\m\n\rho\s}\rp \\
\times\ls\mathcal{P}(z_1,z_2)\lp D\e_1^T\e_2D\rp_{\n\s}-\mathcal{P}(z_1,\zbar_2)\lp D\e_1^TD\e_2^T\rp_{\n\s}\right.\\
\left.-\mathcal{P}(\zbar_1,z_2)\lp\e_1D\e_2D\rp_{\n\s}+\mathcal{P}(\zbar_1,\zbar_2)\lp\e_1\e_2^T\rp_{\n\s}\rs \\
\times\ls -\mathcal{P}(z_1,z_2)\lp p_1p_2\rp\lp Dp_1\rp_\m\lp Dp_2\rp_\rho+\mathcal{P}(z_1,\zbar_2)\lp p_1Dp_2\rp\lp Dp_1\rp_\m p_{2\,\rho}\right.\\
\left.+\mathcal{P}(\zbar_1,z_2)\lp p_1Dp_2\rp p_{1\,\m}\lp Dp_2\rp_\rho-\mathcal{P}(\zbar_1,\zbar_2)\lp p_1p_2\rp p_{1\,\m}p_{2\,\rho}\rs,
\end{multline}
where we have defined
\be
\mathcal{K}=\prod_{i=0}^2\lp z_i-\zbar_i\rp^{p_iDp_i}\prod_{i<j}\left|z_i-z_j\right|^{2p_ip_j}\left|z_i-\zbar_j\right|^{2p_iDp_j},
\ee
from the contraction of the exponentials with each other.

Mapping to the disc, including the integration, and applying our class A selection rules (\ref{eq:ClassASelectionRules}), we find
\begin{multline}
\label{eq:EFCtPPC}
\frac{1}{256}\Tr\lp P_-\slashed{C}M\G^{ab\m\n}\rp\int_0^{2\pi}d\th\lp e^{i\th}-e^{-i\th}\rp^2\int_0^1r_1dr_1\int_0^1r_2dr_2\mathcal{K} \\
\times\ls\frac{\lp\e_1\e_2^T\rp_{\m\n}}{\left|r_1-r_2e^{i\th}\right|^2}-\frac{\lp\e_1D\e_2D\rp_{\m\n}}{\left|1-r_1r_2e^{i\th}\right|^2}\rs\ls\frac{\lp p_1p_2\rp}{\left|r_1-r_2e^{i\th}\right|^2}-\frac{\lp p_1Dp_2\rp}{\left|1-r_1r_2e^{i\th}\right|^2}\rs p_{1\,a}p_{2\,b},
\end{multline}
where now
\be
\mathcal{K}=i^{p_0Dp_0+p_1Dp_1+p_2Dp_2}\left|r_1-r_2e^{i\th}\right|^{2p_1p_2}\left|1-r_1r_2e^{i\th}\right|^{2p_1Dp_2}\prod_{i=1}^2\lp 1-r_i^2\rp^{p_iDp_i}r_i^{2p_0p_i}.
\ee

If we now turn to the contribution from four fermion terms, we find
\begin{multline}
\frac{1}{256}\Tr\lp P_-\slashed{C}M\G^{\m\n\rho\s}\rp\int_0^{2\pi}d\th\int_0^1dr_1\int_0^1dr_2\mathcal{K} \\
\times\ls\frac{e^{i\th}}{\lp r_1-r_2e^{i\th}\rp^2}\lp D\e_1^T\e_2D\rp_{\n\s}\lp Dp_1\rp_\m\lp Dp_2\rp_\rho+\frac{e^{-i\th}}{\lp 1-r_1r_2e^{-i\th}\rp^2}\lp D\e_1^TD\e_2^T\rp_{\n\s}\lp Dp_1\rp_\m p_{2\,\rho}\right. \\
\left.+\frac{e^{i\th}}{\lp 1-r_1r_2e^{i\th}\rp^2}\lp\e_1D\e_2D\rp p_{1\,\m}\lp Dp_2\rp_\rho+\frac{e^{-i\th}}{\lp r_1-r_2e^{-i\th}\rp^2}\lp\e_1\e_2^T\rp p_{1\,\m}p_{2\,\rho}\rs.
\end{multline}
Unlike the eight fermion contribution, this result carries only two explicit factors of momentum.  However, we can write the quantity inside the square brackets as a total derivative with respect to $\th$ and then integrate by parts.  Since
\be
\frac{\pa}{\pa\th}\mathcal{K}=-i\mathcal{K}r_1r_2\lp e^{i\th}-e^{-i\th}\rp\ls\frac{\lp p_1p_2\rp}{\left|r_1-r_2e^{i\th}\right|^2}+\frac{\lp p_1Dp_2\rp}{\left|1-r_1r_2e^{i\th}\right|^2}\rs,
\ee
we have a total contribution
\begin{multline}
\frac{\Tr\lp P_-\slashed{C}M\G^{\m\n\rho\s}\rp}{256}\int_0^{2\pi}d\th\int_0^1dr_1\int_0^1dr_2\mathcal{K}\lp e^{i\th}-e^{-i\th}\rp\ls\frac{\lp p_1p_2\rp}{\left|r_1-r_2e^{i\th}\right|^2}+\frac{\lp p_1Dp_2\rp}{\left|1-r_1r_2e^{i\th}\right|^2}\rs \\
\times\ls\frac{r_1}{r_1-r_2e^{i\th}}\lp D\e_1^T\e_2D\rp_{\n\s}\lp Dp_1\rp_\m\lp Dp_2\rp_\rho-\frac{1}{1-r_1r_2e^{-i\th}}\lp D\e_1^TD\e_2^T\rp_{\n\s}\lp Dp_1\rp_\m p_{2\,\rho}\right. \\
\left.+\frac{1}{1-r_1r_2e^{i\th}}\lp\e_1D\e_2D\rp p_{1\,\m}\lp Dp_2\rp_\rho-\frac{r_1}{r_1-r_2e^{-i\th}}\lp\e_1\e_2^T\rp p_{1\,\m}p_{2\,\rho}\rs.
\end{multline}
Finally, applying our selection rules we are left with
\begin{multline}
\label{eq:FFCtPPC}
-\frac{1}{256}\Tr\lp P_-\slashed{C}M\G^{ab\m\n}\rp\int_0^{2\pi}d\th\lp e^{i\th}-e^{-i\th}\rp^2\int_0^1r_1dr_1\int_0^1r_2dr_2\mathcal{K}\\
\times\ls\frac{\lp\e_1\e_2^T\rp_{\m\n}}{\left|r_1-r_2e^{i\th}\right|^2}+\frac{\lp\e_1D\e_2D\rp_{\m\n}}{\left|1-r_1r_2e^{i\th}\right|^2}\rs\ls\frac{\lp p_1p_2\rp}{\left|r_1-r_2e^{i\th}\right|^2}+\frac{\lp p_1Dp_2\rp}{\left|1-r_1r_2e^{i\th}\right|^2}\rs p_{1\,a}p_{2\,b}.
\end{multline}

The sum of (\ref{eq:EFCtPPC}) and (\ref{eq:FFCtPPC}) give the total result for the terms under consideration,
\be
\label{eq:ClassAPenultimate}
-\frac{1}{128}\Tr\lp P_-\slashed{C}M\G^{ab\m\n}\rp\ls\lp p_1Dp_2\rp\lp\e_1\e_2^T\rp_{\m\n}+\lp p_1p_2\rp\lp\e_1D\e_2D\rp_{\m\n}\rs p_{1\,a}p_{2\,b}\times I_0,
\ee
where (we evaluate the integral in Appendix \ref{app:I0})
\be
I_0=\int_0^{2\pi}d\th\int_0^1r_1dr_1\int_0^1r_2dr_2\frac{\lp e^{i\th}-e^{-i\th}\rp^2}{\left|r_1-r_2e^{i\th}\right|^2\left|1-r_1r_2e^{i\th}\right|^2}\mathcal{K}=-\frac{\pi^3}{3}+\mathcal{O}\lp p^2\rp.
\ee
It actually turns out that we do not need the precise value of this integral, since we ignore an overall normalization and it's the same integral that shows up in all the amplitudes that we eventually compute, but we do need to be sure that the integral is convergent in the limit of zero momentum, i.e. when we set $\mathcal{K}=1$.  Incidentally, this is precisely where trouble appears in more general couplings.  Other terms can produce integrals which do not converge if we replace $\mathcal{K}$ by $1$, but rather require us to keep $\mathcal{K}$ and analytically continue to particular regions of momenta.  The answers they give involve poles (for instance rational functions in which the numerator and denominator are the same order in momenta).  It is also these terms which give disagreeing answers when computed in different pictures.

Putting everything together, the full amplitude for the class A couplings is
\be
\label{eq:ClassAFullAmplitude}
\frac{\pi^3\lp\a'\rp^2}{1536}\Tr\lp P_-\slashed{C}M\G^{ab\m\n}\rp\ls\lp p_1Dp_2\rp\lp\e_1\e_2^T\rp_{\m\n}+\lp p_1p_2\rp\lp\e_1D\e_2D\rp_{\m\n}\rs p_{1\,a}p_{2\,b}.
\ee
Using (\ref{eq:TraceEvaluations}) and (\ref{eq:L(p+1)}), we can compare with (\ref{eq:FixedCouplings}) and find precise agreement up to the overall normalization factor.  Note that when converting from the D-brane action to the three-point amplitude, the terms coupling to $C^{(p-3)}$ and $C^{(p+1)}$ get an extra factor of two relative to the $C^{(p-1)}$ couplings, due to the symmetry between the two NS-NS fields.  This extra factor of two cancels the factor of $1/2$ in these couplings, so that all three can be written in the same form (\ref{eq:ClassAFullAmplitude}).

\subsubsection{Class B couplings}

We turn now to the class B couplings in which each polarization contracts with a momentum.  A priori, there are four-, six-, and eight-fermion terms which can contribute, but it turns out that our selection rules cause the six-fermion contributions to vanish.  From the four-fermion terms we find an expression that reduces to
\begin{multline}
\frac{1}{256}\Tr\lp P_-\slashed{C}M\G^{ab\m\n}\rp p_{1\,a}p_{2\,b}\int_0^{2\pi}\lp e^{i\th}-e^{-i\th}\rp^2d\th\int_0^1r_1dr_1\int_0^1r_2dr_2\mathcal{K} \\
\times\ls\frac{\lp\e_1p_2\rp_\m}{\left|r_1-r_2e^{i\th}\right|^2}+\frac{\lp\e_1Dp_2\rp_\m}{\left|1-r_1r_2e^{i\th}\right|^2}\rs\ls\frac{\lp\e_2p_1\rp_\n}{\left|r_1-r_2e^{i\th}\right|^2}+\frac{\lp\e_2Dp_1\rp_\n}{\left|1-r_1r_2e^{i\th}\right|^2}\rs,
\end{multline}
while from the eight-fermion contribution we find
\begin{multline}
-\frac{1}{256}\Tr\lp P_-\slashed{C}M\G^{ab\m\n}\rp p_{1\,a}p_{2\,b}\int_0^{2\pi}\lp e^{i\th}-e^{-i\th}\rp^2d\th\int_0^1r_1dr_1\int_0^1r_2dr_2\mathcal{K} \\
\times\ls\frac{\lp\e_1p_2\rp_\m}{\left|r_1-r_2e^{i\th}\right|^2}-\frac{\lp\e_1Dp_2\rp_\m}{\left|1-r_1r_2e^{i\th}\right|^2}\rs\ls\frac{\lp\e_2p_1\rp_\n}{\left|r_1-r_2e^{i\th}\right|^2}-\frac{\lp\e_2Dp_1\rp_\n}{\left|1-r_1r_2e^{i\th}\right|^2}\rs.
\end{multline}
In deriving these expressions we have made use of momentum conservation (\ref{eq:MomentumConservation}) and the physical state conditions (\ref{eq:PhysicalStateConditions}), as well as the selection rules mentioned above.  Note however, that no integrations by parts were necessary, since all terms are already explicitly fourth order in momenta.

Thus, all together we find
\begin{multline}
\label{eq:ClassBComputation}
\frac{1}{128}\Tr\lp P_-\slashed{C}M\G^{ab\m\n}\rp\ls\lp\e_1p_2\rp_\m\lp\e_2Dp_1\rp_\n+\lp\e_1Dp_2\rp_\m\lp\e_2p_1\rp_\n\rs p_{1\,a}p_{2\,b}\times I_0 \\
=-\frac{\pi^3\lp\a'\rp^2}{1536}\Tr\lp P_-\slashed{C}M\G^{ab\m\n}\rp\ls\lp\e_1p_2\rp_\m\lp\e_2Dp_1\rp_\n+\lp\e_1Dp_2\rp_\m\lp\e_2p_1\rp_\n\rs p_{1\,a}p_{2\,b},
\end{multline}
where $I_0$ is the same integral we had before.  One may then verify that this expression matches the results derived using T-duality.

\subsection{Results in the $(-1/2,-1/2)\times(-1,0)\times(0,0)$ picture}
\label{subsec:SymmetricPicture}

Now we shall briefly sketch how one can compute the results for our favorite couplings in a different picture and verify that they are independent of how we distribute the picture charge.

The new vertex operators which are required are
\be
V_1^{(-1,0)}(z_1,\zbar_1)=\lp\e_1D\rp_{\m\n}e^{-\phi}\psi^\m e^{ip_1X}(z_1)\lp\pa X^\n+i\lp p_1D\psi\rp\psi^\n\rp e^{ip_1DX}(\zbar_1),
\ee
and
\be
V^{(-1/2,-1/2)}(z_0,\zbar_0)=\lp\mathcal{C}P_+\slashed{F}\rp^{AB}e^{-\hlf\phi(z_0)}e^{-\hlf\phi(\zbar_0)}S_A(z_0)\widetilde{S}_B(\zbar_0)e^{ip_0X(z_0)+ip_0DX(\zbar_0)},
\ee
where
\be
P_+\slashed{F}=P_+\frac{1}{\lp n+1\rp !}F^{(n+1)}_{\m_0\cdots\m_n}\G^{\m_0\cdots\m_n}=P_+\frac{i}{n!}p_{0\,\m_0}C^{(n)}_{\m_1\cdots\m_n}\G^{\m_0\cdots\m_n}=i\slashed{p_0}P_-\slashed{C}.
\ee
In the last step of the above expression we have made use of the physical state conditions for the R-R field.  For the components of $p_0$ which lie along the brane, we can then use momentum conservation to rewrite
\be
p_{0\,a}=-p_{1\,a}-p_{2\,a},
\ee
in order to compare to the results obtained earlier.

One can verify that only fermionic sector expectation values with precisely three gamma matrices can contribute to the couplings of (\ref{eq:FixedCouplings}), and all the non-vanishing results come only from seven- and three-fermion terms.

For class A couplings, adding together the seven- and three-fermion terms (the three-fermion terms require an integration by parts in order to make them explicitly fourth order in momenta) and imposing our selection rules gives a result
\begin{multline}
2^{-\frac{15}{2}}\Tr\lp P_-\slashed{C}M\G^{ab\m\n}\rp p_{0\,a}p_{2\,b}\int_0^{2\pi}d\th\lp e^{i\th}-e^{-i\th}\rp\int_0^1dr_1\int_0^1dr_2\mathcal{K} \\
\times\left\{-\frac{\lp p_1p_2\rp\lp\e_1\e_2^T\rp_{\m\n}}{\left|r_1-r_2e^{i\th}\right|^2}+\frac{1-2r_1^2-r_1^2r_2^2+r_1r_2\lp 3e^{i\th}-e^{-i\th}\rp}{\left|r_1-r_2e^{i\th}\right|^2\left|1-r_1r_2e^{i\th}\right|^2}\lp p_1Dp_2\rp\lp\e_1\e_2^T\rp_{\m\n}\right.\\
\left.+\frac{2-r_1^2+r_2^2+r_1r_2\lp e^{i\th}-3e^{-i\th}\rp}{\left|r_1-r_2e^{i\th}\right|^2\left|1-r_1r_2e^{i\th}\right|^2}\lp p_1p_2\rp\lp\e_1D\e_2D\rp_{\m\n}+\frac{\lp p_1Dp_2\rp\lp\e_1D\e_2D\rp_{\m\n}}{\left|1-r_1r_2e^{i\th}\right|^2}\right\}.
\end{multline}
This appears at first to be much more complicated than the previous cases which involved the integral $I_0$.  However, any odd function of $\th$ will integrate to zero, thus we can symmetrize the integrand with respect to $\th\leftrightarrow -\th$ and find that the expression above is equivalent to
\be
2^{-\frac{13}{2}}\Tr\lp P_-\slashed{C}M\G^{ab\m\n}\rp p_{0\,a}p_{2\,b}\ls\lp p_1Dp_2\rp\lp\e_1\e_2^T\rp_{\m\n}+\lp p_1p_2\rp\lp\e_1D\e_2D\rp_{\m\n}\rs\times I_0.
\ee
If we now make the substitution $p_{0\,a}=-p_{1\,a}-p_{2\,a}$, we see that this expression precisely agrees with (\ref{eq:ClassAPenultimate}), up to an overall factor of $\sqrt{2}$.

Similarly, the class B couplings lead to
\begin{multline}
2^{-\frac{15}{2}}\Tr\lp P_-\slashed{C}M\G^{ab\m\n}\rp p_{0\,a}p_{2\,b}\int_0^{2\pi}d\th\lp e^{i\th}-e^{-i\th}\rp\int_0^1\frac{dr_1}{1-r_1^2}\int_0^1dr_2\mathcal{K} \\
\times\left\{\frac{1+r_1^2}{\left|r_1-r_2e^{i\th}\right|^2}\lp\e_1p_2\rp_\m\lp\e_2p_1\rp_\n+\frac{1+r_1^2}{\left|1-r_1r_2e^{i\th}\right|^2}\lp\e_1Dp_2\rp_\m\lp\e_2Dp_1\rp_\n \right.\\
\left.+\frac{-1+3r_1^2+r_1^2r_2^2-r_1^4r_2^2-r_1r_2\lp 3e^{i\th}-e^{-i\th}\rp+r_1^3r_2\lp e^{i\th}-3e^{-i\th}\rp}{\left|r_1-r_2e^{i\th}\right|^2\left|1-r_1r_2e^{i\th}\right|^2}\lp\e_1p_2\rp_\m\lp\e_2Dp_1\rp_\n\right. \\
\left.+\frac{3r_1^2-r_2^2-r_1^4+3r_1^2r_2^2-r_1r_2\lp 3e^{i\th}-e^{-i\th}\rp+r_1^3r_2\lp e^{i\th}-3e^{-i\th}\rp}{\left|r_1-r_2e^{i\th}\right|^2\left|1-r_1r_2e^{i\th}\right|^2}\lp\e_1Dp_2\rp_\m\lp\e_2p_1\rp_\n\right\}.
\end{multline}
Again taking the part of the integrand which is symmetric in $\th$, the integral reduces to $I_0$ and the result agrees, up to normalization, with (\ref{eq:ClassBComputation}).

\subsection{Relations to other pictures}

There are several other distributions of picture charge which one can consider, but which can be obtained from the calculations we have already done.  For instance, we can replace the $(-3/2,-1/2)$-picture R-R vertex operator of section \ref{subsec:AsymmetricPicture} by a $(-1/2,-3/2)$-picture operator, but it is easy to check that such a change is trivial - the answers do not change.

Anotherl check is to replace the $(-1,0)$-picture NS-NS vertex operator of section \ref{subsec:SymmetricPicture} by the corresponding $(0,-1)$-picture operator.  At the level of expectation values on the upper half-plane, this can be accomplished by noting that the two amplitudes are related by
conjugating $z_1$ while simultaneously sending $\e_1$ to $D\e_1^TD$ and $p_1$ to $Dp_1$.  This conjugation is equivalent to sending $r_1$ to $r_1^{-1}$ and this means that the integrand of $I_0$ gets sent to
\be
\frac{r_1\left|dr_1\right|}{\left|r_1-r_2e^{i\th}\right|^2\left|1-r_1r_2e^{i\th}\right|^2}\rightarrow\frac{r_1^{-3}\left|dr_1\right|}{\left|r_1^{-1}-r_2e^{i\th}\right|^2\left|1-r_1^{-1}r_2e^{i\th}\right|^2}=\frac{r_1\left|dr_1\right|}{\left|1-r_1r_2e^{i\th}\right|^2\left|r_1-r_2e^{i\th}\right|^2},
\ee
while the region of integration, $0\le r_1\le 1$ in this case, does not change.  At the same time, the selection rules ensure that the expression outside the integral is invariant under the given transformation, for both class A and class B.  So, for these particular couplings this shift in picture does not change the result.  This is not true for general terms in the amplitude.

\section{Discussion and Future Directions}
\label{sec:Discussion}

We have shown that the standard four-derivative correction to the Wess-Zumino term in the D-brane action must be supplemented with several other terms which are required by T-duality.  These terms involve not just $C^{(p-3)}$, but also $C^{(p-1)}$ and $C^{(p+1)}$ with legs transverse to the brane, and involve derivatives of the $B$-field as well as derivatives of the metric.  We confirmed the T-duality prediction of these terms by doing worldsheet computations.

However, what we have done so far is just the first step in a larger research program.  We have not written out a full accounting of the corrections which will appear at this order in derivatives.  Invariance under R-R gauge transformations, $B$-field gauge transformations, and spacetime diffeomorphisms will require the presence of additional terms.  Also, there will of course be terms that are higher (or lower) order in the NS-NS fluctuations, and the full collection of terms should be expressible in some form more elegant than (\ref{eq:FixedCouplings}).  Improving our analysis of the implications of spacetime dualities and symmetries should help us move towards a more complete understanding of the Wess-Zumino action at this order.

On the other hand, it is wise to supplement the spacetime analysis with worldsheet computations (perhaps using the techniques of~\cite{Stieberger:2009hq}), as we have done in this paper.  To progress on to more general couplings, we will need to face the unphysical poles that we were able to avoid in the current work.  This requires a careful treatment of worldsheet boundary contributions to our amplitudes, and their relationship to spurious states as in~\cite{Green:1987qu}.  This is the subject of ongoing analysis.  If we then want to get results at higher order in the fluctuations we shall have to compute amplitudes which involve more insertions of both closed and open string operators (see \cite{Fotopoulos:2001pt,Fotopoulos:2002wy} for some examples in the DBI part of the action).

One other technique which may be useful in computing these types of couplings is the use of anomaly inflow arguments as in~\cite{Green:1996dd,Morales:1998ux,Scrucca:1999uz}.  Careful analysis of anomaly cancellation on brane intersections in the presence of varying $B$-fields and R-R fields, as well as a curved background metric might allow us to deduce some of the necessary couplings, and in particular might be useful in providing more of a global picture than the other approaches.

Once we have succeeded in this multi-pronged attack to find a more complete set of four-derivative couplings on the brane worldvolume, the next step will be to understand the physical implications.  In particular, we would like to understand whether these new terms contribute to the global tadpole constraints of flux compactifications (they will surely contribute to the local tadpole equations).  If they did contribute - i.e. if there were contexts in which these extra terms were topologically non-trivial - then it could in principle require a re-evaluation of the consistency of whole classes of vacua.  There could be classes of purported solutions which were no longer valid, or there could be new classes of solutions where previously none were allowed.  It's also possible that the new terms are guaranteed to be globally trivial, and do no more than locally perturb the known set of solutions.  A careful study of the implications of these couplings should shed light on these issues.

Finally, there are other contexts where related higher-derivative couplings may be relevant.  An obvious example is the coupling of the R-R potential to orientifold planes, which is known to receive corrections along the lines of (\ref{eq:OriginalWZAction}), and so should also receive corrections related by T-duality.  The spacetime analysis of this case should be exactly the same as for D-branes, but it would be good to confirm this with another set of worldsheet computations.  Similarly, dualities should map some of these examples to heterotic backgrounds, or to M-theory compactifications, or they may map the D-branes onto other extended objects like NS5-branes.

\acknowledgments

The authors would like to thank Melanie Becker, Aaron Bergman, Rob
Myers, Sav Sethi, and Kelly Stelle for useful discussions and
comments.  We would especially like to thank Mohammad Garousi for
discussions and collaboration in the early stages of this work. This
research was supported in part by NSF Grant No. PHY05-55575, NSF
Grant No. PHY09-06222, NSF Gant No. PHY05-51164, Focused Research
Grant DMS-0854930, Texas A{\&}M University, and the Mitchell
Institute for Fundamental Physics and Astronomy.

\appendix

\section{Evaluation of $I_0$}
\label{app:I0}

We wish to evaluate the integral
\be
I_0=\int_0^{2\pi}d\th\int_0^1r_1dr_1\int_0^1r_2dr_2\frac{\lp e^{i\th}-e^{-i\th}\rp^2}{\left|r_1-r_2e^{i\th}\right|^2\left|1-r_1r_2e^{i\th}\right|^2}\mathcal{K},
\ee
at lowest order in momenta.  The result is known (see, e.g.~\cite{Stefanski:1998yx}), but for completeness we will present our own derivation.  At this order we can set $\mathcal{K}=1$, provided the remaining integral converges.  If we split the integral up into two regions, $r_1\le r_2$ and $r_1\ge r_2$, then we can expand the factors in the denominator of the integrand as Taylor series,
\begin{multline}
I_0=\int_0^{2\pi}d\th\lp e^{i\th}-e^{-i\th}\rp^2\sum_{m_1,n_1,m_2,n_2=0}^\infty\left\{\int_0^1\frac{dr_1}{r_1}\int_0^{r_1}r_2dr_2\lp\frac{r_2}{r_1}\rp^{n_1+n_2}\lp r_1r_2\rp^{m_1+m_2}\right. \\
\left.+\int_0^1\frac{dr_2}{r_2}\int_0^{r_2}r_1dr_1\lp\frac{r_1}{r_2}\rp^{n_1+n_2}\lp r_1r_2\rp^{m_1+m_2}\right\}e^{i\lp n_1-n_2+m_1-m_2\rp\th}.
\end{multline}

The two regions clearly give identical contributions.  Let's now rewrite the sums using $N=n_1+n_2$, $n=(n_1-n_2)/2$, $M=m_1+m_2$, and $m=(m_1-m_2)/2$,
\be
I_0=2\int_0^1\frac{dr_1}{r_1}\int_0^{r_1}r_2dr_2\sum_{N,M=0}^\infty r_1^{M-N}r_2^{M+N}\int_0^{2\pi}d\th\sum_{n=-N/2}^{N/2}\sum_{m=-M/2}^{M/2}\lp e^{i\th}-e^{-i\th}\rp^2e^{2i\lp m+n\rp\th}.
\ee
Note that the angular integral will give a non-zero result if and only if $M$ and $N$ have the same parity (either both even or both odd).  Consider the angular integral at fixed $N$ and $M$.  If $N<M$, then for each allowed value of $n$ there is precisely one allowed $m$ satisfying each $m=-n-1$, $m=-n$, and $m=-n+1$.  Thus, when we expand $(e^{i\th}-e^{-i\th})^2$ and perform the angular integral, the three terms precisely cancel out.  Similarly, the angular integral for $N>M$ gives a vanishing result.  This leaves us only with the case $N=M$,
\beq
I_0 &=& 2\int_0^1\frac{dr_1}{r_1}\int_0^{r_1}r_2dr_2\sum_{N=0}^\infty r_2^{2N}\int_0^{2\pi}d\th\sum_{n,m=-N/2}^{N/2}\lp e^{2i\th}-2+e^{-2i\th}\rp e^{2i\lp m+n\rp\th}\non\\
&=& 4\pi\int_0^1\frac{dr_1}{r_1}\int_0^{r_1}r_2dr_2\sum_{N=0}^\infty\lp N-2\lp N+1\rp+N\rp r_2^{2N}\non\\
&=& -8\pi\int_0^1dr_1\sum_{N=0}^\infty\frac{r_1^{2N+1}}{2N+2}=-2\pi\sum_{N=0}^\infty\frac{1}{(N+1)^2}=-\frac{\pi^3}{3}.
\eeq

\bibliographystyle{utphys}

\begin{thebibliography}{10}

\bibitem{Bershadsky:1995}
M.~Bershadsky, V.~Sadov, C.~Vafa, ``{D-Branes and Topological Field
Theories},''
  \href{http://dx.doi.org/10.1016/0550-3213(96)00026-0}{{\em Nucl. Phys.} {\bf
  B463} (1996)  420--434},
\href{http://arxiv.org/abs/hep-th/9511222}{{\tt
arXiv:hep-th/9511222}}.


\bibitem{Green:1996dd}
M.~B. Green, J.~A. Harvey, and G.~W. Moore, ``{I-brane inflow and anomalous
  couplings on D-branes},''
  \href{http://dx.doi.org/10.1088/0264-9381/14/1/008}{{\em Class. Quant. Grav.}
  {\bf 14} (1997)  47--52},
\href{http://arxiv.org/abs/hep-th/9605033}{{\tt arXiv:hep-th/9605033}}.

\bibitem{Cheung:1997az}
Y.-K.~E. Cheung and Z.~Yin, ``{Anomalies, branes, and currents},''
  \href{http://dx.doi.org/10.1016/S0550-3213(98)00115-1}{{\em Nucl. Phys.} {\bf
  B517} (1998)  69--91},
\href{http://arxiv.org/abs/hep-th/9710206}{{\tt arXiv:hep-th/9710206}}.

\bibitem{Dsagupta:1997}
K.~Dasgupta, D.~P. Jatkar, S.~Mukhi, ``{Gravitational couplings and
Z(2) orientifolds},''
  \href{http://dx.doi.org/10.1016/S0550-3213(98)00155-2}{{\em Nucl. Phys.} {\bf
  B523} (1998)  465--484},
\href{http://arxiv.org/abs/hep-th/9707224}{{\tt
arXiv:hep-th/9707224}}.

\bibitem{Minasian:1997mm}
R.~Minasian and G.~W. Moore, ``{K-theory and Ramond-Ramond charge},'' {\em
  JHEP} {\bf 11} (1997)  002,
\href{http://arxiv.org/abs/hep-th/9710230}{{\tt arXiv:hep-th/9710230}}.


\bibitem{sethitalks}
S.~Sethi , ``{Fluxes, Geometries and Non-Geometries},'' talks at
BIRS workshop (March 2010) and Strings 2010 (March 2010).

\bibitem{mcoristsethitoappear}
J.~McOrist and S.~Sethi to appear.

\bibitem{Buscher:1987sk}
T.~H. Buscher, ``{A symmetry of the string background field equations},''
\href{http://dx.doi.org/10.1016/0370-2693(87)90769-6}{{\em Phys. Lett.} {\bf
  B194} (1987)  59}.

\bibitem{Bergshoeff:1995as}
E.~Bergshoeff, C.~M. Hull, and T.~Ortin, ``{Duality in the type II superstring
  effective action},''
  \href{http://dx.doi.org/10.1016/0550-3213(95)00367-2}{{\em Nucl. Phys.} {\bf
  B451} (1995)  547--578},
\href{http://arxiv.org/abs/hep-th/9504081}{{\tt arXiv:hep-th/9504081}}.

\bibitem{Hohm:2010jy}
O.~Hohm, C.~Hull, and B.~Zwiebach, ``{Background independent action for double
  field theory},''
\href{http://arxiv.org/abs/1003.5027}{{\tt arXiv:1003.5027 [hep-th]}}.

\bibitem{inprogress}
K.~Becker, G.~Guo, and D.~Robbins work in progress.

\bibitem{Bachas:1999um}
C.~P. Bachas, P.~Bain, and M.~B. Green, ``{Curvature terms in D-brane actions
  and their M-theory origin},'' {\em JHEP} {\bf 05} (1999)  011,
\href{http://arxiv.org/abs/hep-th/9903210}{{\tt arXiv:hep-th/9903210}}.

\bibitem{Eisenhart}
L.~P. Eisenhart, {\em Riemannian Geometry}.
\newblock Princeton University Press, 1950.

\bibitem{Shelton:2005cf}
J.~Shelton, W.~Taylor, and B.~Wecht, ``{Nongeometric flux compactifications},''
  {\em JHEP} {\bf 10} (2005)  085,
\href{http://arxiv.org/abs/hep-th/0508133}{{\tt arXiv:hep-th/0508133}}.

\bibitem{Garousi:2010ki}
M.~R. Garousi, ``{Ramond-Ramond field strength couplings on D-branes},''
  \href{http://dx.doi.org/10.1007/JHEP03(2010)126}{{\em JHEP} {\bf 03} (2010)
  126},
\href{http://arxiv.org/abs/1002.0903}{{\tt arXiv:1002.0903 [hep-th]}}.

\bibitem{Green:1987qu}
M.~B. Green and N.~Seiberg, ``{Contact interactions in superstring theory},''
\href{http://dx.doi.org/10.1016/0550-3213(88)90549-4}{{\em Nucl. Phys.} {\bf
  B299} (1988)  559}.

\bibitem{Craps:1998fn}
B.~Craps and F.~Roose, ``{Anomalous D-brane and orientifold couplings from the
  boundary state},''
  \href{http://dx.doi.org/10.1016/S0370-2693(98)01438-5}{{\em Phys. Lett.} {\bf
  B445} (1998)  150--159},
\href{http://arxiv.org/abs/hep-th/9808074}{{\tt arXiv:hep-th/9808074}}.

\bibitem{Craps:2000zr}
B.~Craps, ``{D-branes and boundary states in closed string theories},''
\href{http://arxiv.org/abs/hep-th/0004198}{{\tt arXiv:hep-th/0004198}}.

\bibitem{Stefanski:1998yx}
B.~Stefanski, Jr., ``{Gravitational couplings of D-branes and O-planes},''
  \href{http://dx.doi.org/10.1016/S0550-3213(99)00147-9}{{\em Nucl. Phys.} {\bf
  B548} (1999)  275--290},
\href{http://arxiv.org/abs/hep-th/9812088}{{\tt arXiv:hep-th/9812088}}.

\bibitem{Morales:1998ux}
J.~F. Morales, C.~A. Scrucca, and M.~Serone, ``{Anomalous couplings for
  D-branes and O-planes},''
  \href{http://dx.doi.org/10.1016/S0550-3213(99)00217-5}{{\em Nucl. Phys.} {\bf
  B552} (1999)  291--315},
\href{http://arxiv.org/abs/hep-th/9812071}{{\tt arXiv:hep-th/9812071}}.

\bibitem{Garousi:1996ad}
M.~R. Garousi and R.~C. Myers, ``{Superstring scattering from D-branes},''
  \href{http://dx.doi.org/10.1016/0550-3213(96)00316-1}{{\em Nucl. Phys.} {\bf
  B475} (1996)  193--224},
\href{http://arxiv.org/abs/hep-th/9603194}{{\tt arXiv:hep-th/9603194}}.

\bibitem{Gubser:1996wt}
S.~S. Gubser, A.~Hashimoto, I.~R. Klebanov, and J.~M. Maldacena,
  ``{Gravitational lensing by $p$-branes},''
  \href{http://dx.doi.org/10.1016/0550-3213(96)00182-4}{{\em Nucl. Phys.} {\bf
  B472} (1996)  231--248},
\href{http://arxiv.org/abs/hep-th/9601057}{{\tt arXiv:hep-th/9601057}}.

\bibitem{Liu:2001qa}
H.~Liu and J.~Michelson, ``{*-trek III: The search for Ramond-Ramond
  couplings},'' \href{http://dx.doi.org/10.1016/S0550-3213(01)00403-5}{{\em
  Nucl. Phys.} {\bf B614} (2001)  330--366},
\href{http://arxiv.org/abs/hep-th/0107172}{{\tt arXiv:hep-th/0107172}}.

\bibitem{Billo:1998vr}
M.~Billo {\em et al.}, ``{Microscopic string analysis of the D0-D8 brane system
  and dual R-R states},''
  \href{http://dx.doi.org/10.1016/S0550-3213(98)00296-X}{{\em Nucl. Phys.} {\bf
  B526} (1998)  199--228},
\href{http://arxiv.org/abs/hep-th/9802088}{{\tt arXiv:hep-th/9802088}}.

\bibitem{Stieberger:2009hq}
S.~Stieberger, ``{Open \& closed vs. pure open string disk
amplitudes},'' \href{http://arxiv.org/abs/0907.2211}{{\tt
arXiv:0907.2211 [hep-th]}}.

\bibitem{Fotopoulos:2001pt}
A.~Fotopoulos, ``{On (alpha')**2 corrections to the D-brane action for non-
  geodesic world-volume embeddings}",
  {\em JHEP} {\bf 09} (2001)  005,
\href{http://arxiv.org/abs/hep-th/0104146}{{\tt arXiv:hep-th/0104146}}.

\bibitem{Fotopoulos:2002wy}
A.~Fotopoulos and A.~Tseytlin, ``{On gravitational couplings in D-brane action}",
  {\em JHEP} {\bf 12} (2002)  001,
\href{http://arxiv.org/abs/hep-th/0211101}{{\tt arXiv:hep-th/0211101}}.

\bibitem{Scrucca:1999uz}
C.~A. Scrucca and M.~Serone, ``{Anomalies and inflow on D-branes and
  O-planes},'' \href{http://dx.doi.org/10.1016/S0550-3213(99)00357-0}{{\em
  Nucl. Phys.} {\bf B556} (1999)  197--221},
\href{http://arxiv.org/abs/hep-th/9903145}{{\tt arXiv:hep-th/9903145}}.

\end{thebibliography}

\providecommand{\href}[2]{#2}\begingroup\raggedright\endgroup


\end{document}